\documentclass[aps,prl,superscriptaddress,twocolumn]{revtex4-2}
\usepackage{amsmath,bm}
\usepackage{graphicx,epsfig,braket,amssymb,amsfonts,eufrak,tabularx,multirow,amsmath,graphicx,color}
\usepackage{newtxtext}
\usepackage[colorlinks=true]{hyperref}

\newcommand{\bd}{\begin{displaymath}}
\newcommand{\ed}{\end{displaymath}}
\renewcommand{\v}[1]{{\bf #1}}
\newcommand{\bpm}{\begin{pmatrix}}
\newcommand{\epm}{\end{pmatrix}}
\newcommand{\nn}{\nonumber \\}

\begin{document}

\title{Intrinsic Magnon Orbital Hall Effect in Honeycomb Antiferromagnets}

\author{Gyungchoon Go}
\affiliation{Department of Physics, Korea Advanced Institute of Science and Technology, Daejeon 34141, Korea}

\author{Daehyeon An}
\affiliation{Department of Physics, Korea Advanced Institute of Science and Technology, Daejeon 34141, Korea}

\author{Hyun-Woo Lee}
\affiliation{Department of Physics, Pohang University of Science and Technology, Pohang 37673, Korea}

\author{Se Kwon Kim}
\affiliation{Department of Physics, Korea Advanced Institute of Science and Technology, Daejeon 34141, Korea}

\begin{abstract}
We theoretically investigate the transport of magnon orbitals in a honeycomb antiferromagnet. We find that the magnon orbital Berry curvature is finite even without spin-orbit coupling and thus the resultant magnon orbital Hall effect is an intrinsic property of the honeycomb antiferromagnet rooted only in the exchange interaction and the lattice structure. Due to the intrinsic nature of the magnon orbital Hall effect, the magnon orbital Nernst conductivity is estimated to be orders of magnitude larger than the predicted values of the magnon spin Nernst conductivity that requires finite spin-orbit coupling. For the experimental detection of the predicted magnon orbital Hall effect, we invoke the magnetoelectric effect that couples the magnon orbital and the electric polarization, which allows us to detect the magnon orbital accumulation through the local voltage measurement. Our results pave a way for a deeper understanding of the topological transport of the magnon orbitals and also its utilization for low-power magnon-based orbitronics, namely magnon orbitronics.
\end{abstract}

\maketitle

\emph{Introduction.}\textemdash
The collective low-energy excitations of the ordered materials are of great interest in condensed matter physics.
One of the representative examples is a quantum of spin waves, called a magnon which is a charge-neutral boson in magnetic materials. Magnons have been intensively studied for technological applications since they can realize Joule-heating-free information transport and processing as well as wave-based computing~\cite{Chumak2015}. In addition, for fundamental interest, various topological properties of magnon bands have been investigated in the context of the magnon Hall effect~\cite{Katsura2010,Han2017,Matsumoto2011,Onose2010,Kim2016,Owerre2016} and the spin Nernst effect~\cite{Cheng2016,Zyuzin2016}.
According to the existing theories, the finite Hall response of magnons can occur as manifiestations of spin-orbit coupling through the Dzyaloshinskii-Moriya interaction (DMI)~\cite{Onose2010, Kim2016, Owerre2016, Cheng2016,Zyuzin2016}
or through the magnon-phonon coupling~\cite{Takahashi2016, Zhang2019, Go2019, Park2020}. The Hall response can also occur in spin texture systems~\cite{Katsura2010, Han2017} with the scalar spin chirality, which acts as an effective spin-orbit coupling.

In electronics systems, on the other hand, there have been studies showing that
electrons can exhibit a Hall effect {\it without} spin-orbit coupling, owing to their orbital degree of freedom ~\cite{Tanaka2008, Kontani2008, Kontani2009, Tanaka2010}.
This discovery evoked a surge of interest in electron-orbital transport phenomena
such as the orbital Hall effect~\cite{Go2018, Jo2018, Sala2022, Bhowal2021, Pezo2022} and the orbital torque~\cite{Go2020PRR, Lee2021, Hayashi2022}. Moreover, there have been
theoretical suggestions that orbital-dependent electron transport critically affects electron spin dynamics
when spin-orbit coupling is present. For instance, it has been suggested~\cite{Kontani2009, Tanaka2010, Go2018} that the orbital Hall effect may play a crucial role in the spin Hall effect.

Motivated by the aforementioned advancement of our understanding of electron orbitals, the orbital motion of magnons has started gathering attention recently in magnetism and spintronics. For example, the circulating magnonic modes have been investigated in confined geometries such as whispering gallery mode cavities~\cite{Haigh2016, Sharma2017, Osada2018}, magnetic nanocylinders and nanotubes~\cite{Jiang2020, GonzalezJMMM2010, LeeSH2021,LeeSH2022}. Also, the orbital magnetization of magnons has been pointed out as the origin of weak ferromagnetism in a noncollinear kagome antiferromagnet (AFM) with the DMI~\cite{Neumann2000}. Recently, the orbital-angular-momentum textures of the magnon bands have been realized in collinear magnets with nontrivial networks of exchange interaction~\cite{Fishman2022,Fishman2022b}. Furthermore, inspired by achievements in topological metamaterials~\cite{Wu2015,He2016,Li2018}, topological magnonic modes carrying the magnon current circulation have been demonstrated in honeycomb magnets with exchange-interaction modulation~\cite{Huang2022}.
Despite the strong interest in magnon orbitals, studies on their transport properties are very limited~\cite{Neumann2000}.
In particular, it is an open question whether the magnon orbital degree of freedom can induce
a Hall phenomenon {\it without} spin-orbit coupling, just as its electron counterpart can.

\begin{figure*}[t!]
\includegraphics[width=1.8\columnwidth]{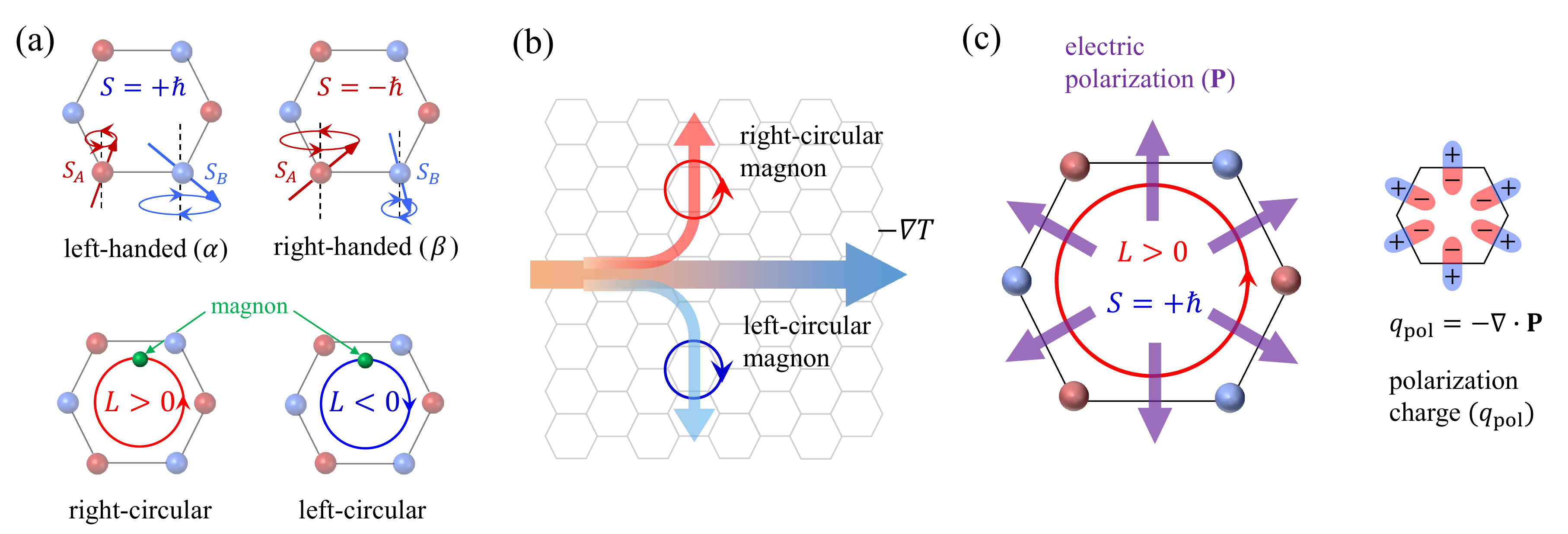}
\caption{(a) Schematics of the magnon spin ($S$) and magnon orbital ($L$) in the honeycomb AFM. The magnon spin and the magnon orbital are determined by, respectively, the precessions of constituent spins within the sites and the intersite hopping. (b) Schematics of the magnon orbital Nernst effect, where a temperature gradient $\boldsymbol{\nabla}T$ induces a net magnon-orbital current in a transverse direction consisting of oppositely-moving right-circular magnons and left-circular magnons. (c) Schematics of the polarization $\mathbf{P}$ induced by the circulating magnonic spin current in the case of orbital $L > 0$ and spin $S = +\hbar$ (left) and the corresponding polarization charge $q_\text{pol} = - \boldsymbol{\nabla} \cdot \mathbf{P}$ (right). The sign of the polarization charge is determined by the product of the sign of the magnon spin $S$ and the direction of the spin current circulation, i.e., the sign of $L$.}\label{fig:1}
\end{figure*}

In this Letter, we answer this question by investigating the transport of magnon orbitals.
For the model system of 2D honeycomb AFMs, we demonstrate
that the magnon orbitals can exhibit a Hall effect, namely the magnon orbital Hall effect, \emph{without} spin-orbit coupling. The magnon orbital moment represents a magnon current circulation as shown in Fig.~\ref{fig:1}(a). We find that application of a longitudinal temperature gradient drives thermal magnons to opposite transverse directions depending on their orbital characters~[see Fig.~\ref{fig:1}(b)], giving rise to a magnon Hall phenomenon, the magnon orbital Nernst effect. The estimated magnitude of the magnon orbital Nernst conductivity is orders-of-magnitude larger than the known values of the magnon spin Nernst conductivity induced by the Dzyaloshinskii-Moriya interaction~\cite{Cheng2016}, revealing the hitherto unrecognized role of the magnon orbital in magnon transport. To propose an experimental method for detecting the accumulation of the magnon orbital at the sample edges induced by the magnon orbital Hall effect, we invoke the magnetoelectric effect by which a magnonic spin current induces an electric dipole moment. Since the magnon orbital moment can be regarded as a magnonic spin-current circulation, the aforementioned magnetoelectric effect dictates that the magnon orbital moment should  engender a polarization charge in a two-dimensional space~\cite{Katsura2005, Vignale2011, Matsumoto2011}~[see Fig.~\ref{fig:1}(c)] and thus the magnon orbital accumulation should be accompanied by the accumulation of the polarization charge. We estimate the electric voltage profile induced by the magnon orbital moment accumulation, which is shown to be within current experimental reach. Owing to strenuous efforts to realize magnetism in various 2D magnetic crystals, our proposal can be tested in a number of transition metal compounds that are known to host honeycomb AFMs [$e.g.,$ {\it M}P{\it X}$_3$ ({\it M} = Fe, Ni, Mn; {\it X}=S, Se)]~\cite{Wang2016,Kim2019,LEFLEM1982,Jiang2021}.

\begin{figure*}[t!]
\includegraphics[width=1.85\columnwidth]{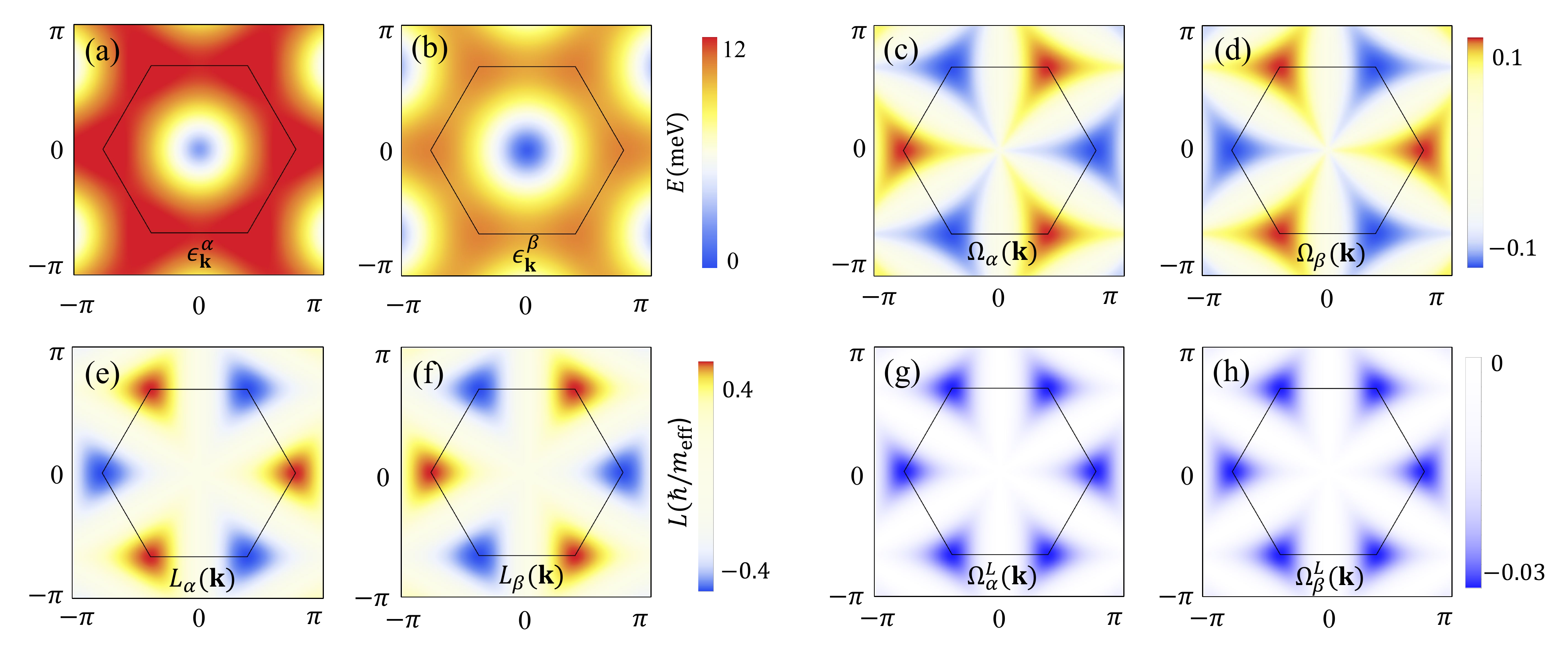}
\caption{ (a, b) Magnon band structures [Eq.~(\ref{eq:band})], (c, d) the Berry curvatures [Eq.~(\ref{eq:Omega})], (e, f) the equilibrium orbital moment structures [Eq.~(\ref{magOAM0})], and (g, h) the orbital Berry curvatures [Eq.~(\ref{OAMBC})] of two magnonic states denoted by $\alpha$ and $\beta$. For material parameters, we take $J = 1.54$ meV and $KS= 0.0086$ meV, and $g \mu_B B/J = 0.25$. For (e) and (f), $m_{\rm eff}$ is the magnon effective mass at the Dirac points.}\label{fig:2}
\end{figure*}

\emph{Model construction.}\textemdash
Here we consider a 2D AFM on a honeycomb lattice
\begin{align}
H = J \sum_{\langle i,j \rangle} \v S_i\cdot \v S_j - K \sum_i (S_{i,z})^2 + g \mu_B B \sum_i S_{i,z} \, ,
\end{align}
where $J( > 0)$ is the antiferromagnetic exchange coupling and $K( > 0)$ is the easy-axis anisotropy,
$g$ is the g-factor, $\mu_B$ is the Bohr magneton, and $B$ is the applied magnetic field.
Note that our model does not include the DMI which comes from spin-orbit coupling and thus does not exhibit the magnon spin Nernst effect~\cite{Kim2016,Owerre2016,Cheng2016,Zyuzin2016}.
In this Letter, we focus on the case where a ground state is the collinear N\'{e}el state along the $z$-axis. Performing the Holstein-Primakoff transformation and taking the Fourier transformation, we have
\begin{align}
H = \frac12 \sum_{\v k} \psi^\dag_\v k {\cal H}_{\v k} \psi_\v k, \qquad \psi_\v k = (a_\v k, b_\v k, a^\dag_{-\v k}, b^\dag_{-\v k})^T \, ,
\end{align}
with the momentum-space Hamiltonian
\begin{align}
{\cal H}_{\v k} = JS\left(
                    \begin{array}{cccc}
                      3 + \kappa_+ & 0 & 0 &  f_{\v k} \\
                      0 & 3 + \kappa_- &  f^\ast_{\v k} & 0 \\
                      0 &  f_{\v k} & 3 + \kappa_+ &  0 \\
                       f^\ast_{\v k} & 0 & 0 & 3 + \kappa_-\\
                    \end{array}
                  \right) \, ,
\end{align}
where $\kappa_\pm = (2K \pm g\mu_B B)/J$ and $f_{\v k} = \sum_{j} e^{i \v k \cdot \v a_{j}}$
with ${\v a}_1 = \frac{a}{2}(\sqrt3,-1)$, ${\v a}_2 = a(0,1)$, and  ${\v a}_3 = -\frac{a}{2}(\sqrt3,1)$. The magnon excitations can be described by the generalized Bogoliubov-de Gennes equation in the particle-hole space representation~\cite{Zhang2019, Li2020}. In this representation, the pseudo-energy-eigenvalue satisfies $\bar\epsilon_{n,\v k} = ({\sigma_3} \epsilon_{\v k})_{nn}$, where ${\sigma_3} = \rm diag(1,1,-1,-1)$ is the Pauli matrix acting on the particle-hole space and $\epsilon(\mathbf{k})$ are the magnon bands given by
\begin{align}
\epsilon_{\v k}^{\alpha/\beta} = \epsilon_{\v k}^0 \pm g\mu_B B S \, ,
\label{eq:band}
\end{align}
where $\epsilon_{\v k}^0 = JS\sqrt{(3 + \kappa)^2 - |f_{\v k}|^2}$ and $\kappa = 2K/J$ (see Supplemental Material for calculation details).
Here, the indices $\alpha$ and $\beta$ stand for two magnonic bands with opposite spin angular momenta~[see Fig~\ref{fig:1}(a)]. The topological property of the magnonic state $|n\rangle$ is characterized by the Berry curvature:
\begin{align}
\Omega_n(\v k) = \frac{\partial A^{n}_y(\v k)}{\partial k_x} - \frac{\partial A^{n}_x(\v k)}{\partial k_y} \, ,
\label{eq:Omega}
\end{align}
where ${\v A}^n({\v k}) = {\langle n|{\sigma_3} i\partial_{\v k}|n\rangle}/{\langle n|{\sigma_3} |n\rangle}$ is the Berry connection. Figures~\ref{fig:2}(a-d) show the magnon band structures $\epsilon^n_\mathbf{k}$ and the corresponding Berry curvatures $\Omega_n (\mathbf{k})$ with $n = \alpha, \beta$.
In the honeycomb AFM, the broken inversion symmetry allows a non-zero Berry curvature without the DMI.
Also, because of the combined symmetry of the time-reversal ($\cal T$) and a $180^\circ$ spin rotation around the $x$-axis (${\cal C}_x$) of the Hamiltonian,
the energy spectra are even in momentum space ($\epsilon_{\v k} = \epsilon_{-\v k}$), whereas the magnon Berry curvatures are odd $[\Omega_n (\v k) = -\Omega_n (-\v k)]$~\cite{Cheng2016}.
Therefore, the momentum-space integration of the magnon Berry curvature weighed by the Bose-Einstein distribution is zero for each band, indicating the absence of the Hall transport of magnons and their spins.

\emph{Magnon orbital Hall effect.}\textemdash
Although the momentum-space integration of the magnon Berry curvatures vanishes, the nonvanishing and $\v k$-odd structure of the Berry curvature opens up a possibility for topological transport of certain quantities. If there is a momentum-dependent quantity whose profile is also odd in $\v k$, its Hall effect can be present. We show below that this is indeed the case for the magnon orbital moment since it holds the same symmetry property in the momentum space as the Berry curvature~\cite{Fishman2022}, i.e., $\v k$-odd in the presence of the ${\cal T C}_x$ symmetry with broken inversion symmetry.

From the orbital moment operator ${\hat {\v L}} = \frac{1}{4}\left(\v r \times \v v - \v v \times \v r\right)$, we read the matrix element of the magnon orbital moment~\cite{Bhowal2021,Pezo2022}
\begin{align}\label{magOAM0}
\langle n|{\hat {\v L}}|m\rangle & = \langle n|\left(\frac{\v r \times \v v - \v v \times \v r}{4}\right)|m\rangle\nn
&= \frac{1}{2\hbar}{\rm Im}\langle \partial_{\v k} n|\times {\cal H}_{\v k}|\partial_{\v k} m\rangle\nn
&\hspace{5mm}-\frac{1}{4 \hbar}(\bar\epsilon_{n,\v k} + \bar\epsilon_{m,\v k}) {\rm Im}\langle \partial_{\v k} n|{\sigma_3}|\partial_{\v k} m\rangle.
\end{align}
By taking the diagonal element of~\eqref{magOAM0}, one recovers the well-known formula of the intrinsic orbital moment~\cite{Xiao2005, Thonhauser2005, Shi2007}.
The magnon orbital moment profiles $L_n(\v k) =\langle n|{\hat L}_{z}|n\rangle$ of the two magnons are shown in Fig.~\ref{fig:2}(e) and (f)~\footnote{Note that the orbital moment $\sim \mathbf{r} \times \mathbf{v}$ differs from the orbital angular momentum $\sim \mathbf{r} \times \mathbf{p}$ by dimension of mass in the electronic case~\cite{Bhowal2021}.}. In agreement with the momentum-space texture of the magnon orbital angular momentum in Ref.~\cite{Fishman2022, Fishman2022b}, the evaluated magnon orbital moment has the $C_3$ rotation symmetry with $\v k$-odd structure.
Because $L^n_z(\v k)$ = $-L^n_z(-\v k)$ and $\epsilon_{\v k} = \epsilon_{-\v k}$, the total magnon orbital moment is zero in equilibrium. However, our system can exhibit the intrinsic magnon orbital Hall transport because both $L^n_z(\v k)$ and $\Omega^n_{z} (\v k)$ are odd in $\v k$,
and thus their product $L^n_z(\v k) \Omega^n_{z} (\v k)$ is $\v k$-even. Analogous to the generalized Berry curvature~\cite{Li2020, Park2020b}, we write the orbital Berry curvature which characterizes the magnon orbital Hall transport as follows:
\begin{align}\label{OAMBC}
\Omega^{{L}}_{n}(\v k) &= \sum_{m \neq n} ({\sigma_3})_{mm}  ({\sigma_3})_{nn}
\frac{2\hbar^2 {\rm Im} \left[\langle n| j^L_{z,y} |m\rangle
\langle m| v_x |n\rangle \right] }{(\bar\epsilon_{n,\v k} - \bar\epsilon_{m,\v k})^2} \, ,
\end{align}
where $j^{L}_{z,y}  = (v_y {\sigma_3}{\hat L}_{z} + {\hat L}_{z} {\sigma_3} v_y) / 4$ is the magnon orbital current operator,
and $v_i = \frac{1}{\hbar}\frac{\partial {\cal H}_{\v k}}{\partial k_i}$ is the velocity operator.
Note the summation is performed in the particle-hole space.
The profiles of the orbital Berry curvatures of the two magnon modes are shown in Fig.~\ref{fig:2}(g) and (h).
As expected, the profiles are even in $\v k$ and thus their momentum-space integrations are finite, indicating the existence of the Hall effect of the magnon orbitals. We emphasize that our model has no spin-orbit coupling term such as the DMI. Also, the orbital Berry curvature remains finite when both $B$ and $K$ approach zero as long as the antiferromagnetic ground state is maintained. Therefore, the magnon orbital Berry curvature and the resultant Hall effect are intrinsic properties of the honeycomb AFM that originated solely from the exchange interaction and the lattice geometry.

\begin{figure}[t]
\includegraphics[width=1.0\columnwidth]{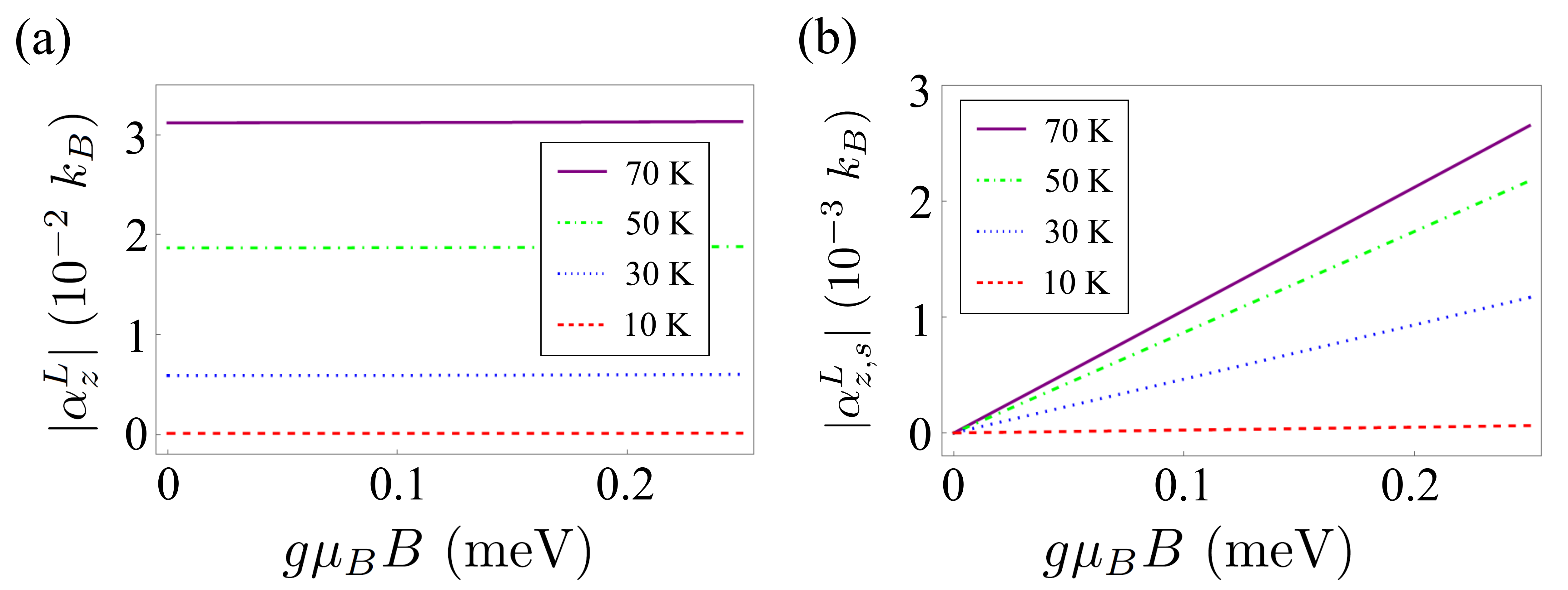}
\caption{(a) Magnetic-field dependence of the magnon orbital Nernst conductivity $\alpha^L_z$ and (b) the spin-polarized component of the magnon orbital Nernst conductivity $\alpha^L_{z,s} = \alpha^L_{z,\alpha} - \alpha^L_{z,\beta}$. See the main text for the details.}\label{fig:3}
\end{figure}

The magnon orbital Berry curvature leads to the transverse magnon orbital current in response to an external perturbation, namely the magnon orbital Hall effect. The linear response equation of the transverse magnon orbital current driven by a temperature gradient is given by $(J^L_{z})_y = -\alpha^{L}_{z} \partial_x T$~\cite{Li2020},
where $\alpha^{L}_{z} = \alpha^{L}_{z,\alpha} + \alpha^{L}_{z,\beta}$ is the magnon orbital Nernst conductivity with
$\alpha^{L}_{z,n} = \frac{2 k_B}{\hbar V} \sum_{\v k} c_1(\rho_n)\Omega^{L}_{n}(\v k)$,
where $k_B$ is the Boltzmann constant, $c_1(\rho) = (1+\rho){\rm ln}(1+\rho) - \rho {\rm ln}\rho$,
and $\rho_n = (e^{\epsilon_n/k_B T}-1)^{-1}$ is the Bose-Einstein distribution.
In Fig.~\ref{fig:3}(a), we show the orbital Nernst conductivity for different temperatures by using the material parameters of MnPS$_3$: $J = 1.54$ meV and $KS= 0.0086$ meV~\cite{Wildes1998}. To compare the orbital Nernst conductivity with the spin Nernst conductivity, we take the constant effective mass approximation at the Dirac points where the Berry curvatures are maximized. The magnon orbital Nernst conductivity is estimated to be about $10^{-2}$${k}_B$. This value is $10^3$ times larger than the estimated values of the magnon spin Nernst conductivity of honeycomb AFMs in the presence of the DMI~\cite{Cheng2016}. This is our main result: The honeycomb AFM exhibits the magnon orbital Hall effect without spin-orbit coupling and, therefore, its magnitude is orders-of-magnitude larger than the magnon spin Nernst conductivity that requires spin-orbit coupling. Note that the orbital Nernst conductivity is almost independent of the magnetic field in Fig.~\ref{fig:3}(a), because the magnon eigenstates are unaffected by the magnetic field.

\emph{Magnon-orbital-induced polarization.}\textemdash
The magnon orbital Hall effect induces the accumulation of the magnon orbital at the edges of a system. To propose an experimental method to detect the magnon orbital accumulation, here we develop a phenomenological model for the transverse voltage profile induced by the longitudinal temperature gradient via the magnon orbital Nernst effect. We emphasize that the following theory is qualitative in nature and thus intended to provide the order-of-magnitude estimation, not quantitative predictions. To begin with, let us review the relation between the spin current and the polarization. The spin current from a noncollinear spin configuration is known to induce an electric polarization in magnetic materials by the combined action of the atomic spin-orbit interaction and the orbital hybridization~\cite{Katsura2005, Vignale2011}: 
\begin{align}\label{inducedP}
{\v P} 
= - \frac{ea}{E_{\rm SO}} {\v e}_{12} \times {\v I}_s,
\end{align}
where $e$ is the magnitude of the electron charge, $a$ is the distance between the two sites, ${\v e}_{12}$ is the unit vector connecting two sites, ${\v I}_s= J ({\v S}_1 \times {\v S}_2)$ is the spin current from site $1$ to site $2$,
and the energy scale $E_{\rm SO}$ is inversely proportional to the spin-orbit coupling strength.
In the ground state of a collinear magnet, there is no spin current $({\v I}_s = 0)$ and thus no electric polarization ($\v P = 0$). However, the magnon  consists of spatially-varying noncollinear deviations from the ground state. Therefore, a magnon current in a collinear magnet gives rise to a finite spin current ${\v I}_s$~\cite{Schutz2003,Schutz2004PRB}. For the magnonic spin current, we can invoke Eq.~\eqref{inducedP} to compute the induced polarization since the characteristic time scale of the magnon is generally much longer than that of the electron hopping process. The spin current carried by a single magnon is given by ${\v I}_s = -S (v/a )\hat{\v z}$, which leads to the electric polarization
\begin{align}\label{inducedP2}
{\v P} = \frac{e S}{E_{\rm SO}} ({\v v} \times \hat{\v z}),
\end{align}
where $S = \pm \hbar$ is the magnon spin and $\v v = v\, {\v e}_{12}$ is the magnon velocity.
By considering the typical energy scale of $E_{\rm SO}$, it has been predicted in Refs.~\cite{Katsura2005, Vignale2011,Xia2023} 
that a measurable electric polarization can be induced by a magnonic spin current in magnetic materials~\footnote{The electric polarization can be also caused by the Lorentz transformation of the magnetic dipole from a moving frame to the laboratory $\v P = \frac{\v v \times \boldsymbol{\mu}}{ c^2} =  \frac{g}{2}\frac{e \hbar}{m_{e} c^2}(\v v \times \hat{\v z})$, where $\v v$ and $\boldsymbol{\mu} (= g \mu_B \v z)$ are the velocity and the magnetic moment of the magnetic dipole, respectively, and $m_{e}$ is the electron mass~\cite{Hirsch1999, Schutz2003, MeierPRL2003}. However, this effect is extremely small because the denominator $m_{e} c^2$ is order of MeV.}.
This magnetoelectric effect allows us to relate the magnon orbital motion
(i.e., the circulating magnonic spin current) and the polarization charge density.
For the qualitative understanding, we schematically depict the electric polarization produced by the magnonic spin-current circulation around a hexagon in a honeycomb lattice in Fig.~\ref{fig:1}(c).
The magnonic spin-current circulation induces the electric polarization pointing outward or inward (and therefore the positive or negative polarization charge density), depending on the product of the magnon-spin sign and the magnon-orbital-moment sign.

Now let us consider the situation where the nonequilibrium accumulation of the magnon orbital moment is generated at the edges of the sample by a temperature gradient via the magnon orbital Nernst effect. In the absence of an external field along the $z$-direction, there would be a finite accumulation of the magnon orbital moment at the edges, but there would be no induced electric polarization, for spin-up magnons and spin-down magnons are equally populated and thus their contributions to the electric polarization cancel each other. However, when we apply an external magnetic field, spin-up magnons and spin-down magnons are populated unequally and thus the net spin density of magnons becomes finite. Consequently, the magnon orbital accumulation and also the magnon orbital Hall current are spin-polarized in the presence of the external field. In particular, the spin-polarized component of the magnon orbital Nernst conductivity $\alpha^{L}_{z,s} (= \alpha^{L}_{z,\alpha} - \alpha^{L}_{z,\beta}$) is zero when the magnetic field is zero and becomes finite as the magnetic field is applied as shown in Fig.~\ref{fig:3}(b).

\begin{figure}[t]
\includegraphics[width=1.0\columnwidth]{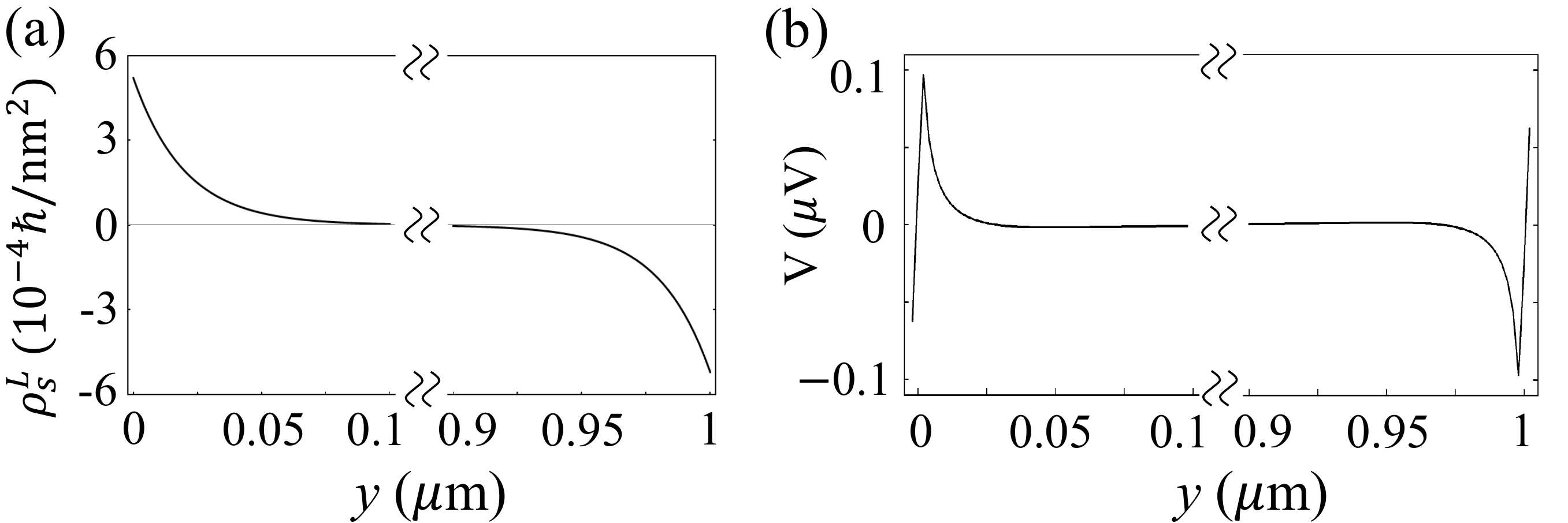}
\caption{(a) The profile of the spin-polarized nonequilibrium magnon orbital accumulation $\rho^L_s$ in the $y$ direction and (b) the corresponding voltage $V$ induced by the nonequilibrium polarization. The considered system size is 1$\mu$m $\times 1\mu$m and a temperature gradient is applied in the $x$ direction. See the main text for the details.}
\label{fig:4}
\end{figure}

Instead of the magnon orbital accumulation, what is directly related to the observable electric polarization is the spin-polarized magnon orbital accumulation $\rho^L_s = \rho^L_\alpha - \rho^L_\beta$. To estimate the spin-polarized magnon orbital accumulation that is induced by the magnon orbital Nernst effect, we use the phenomenological drift-diffusion formalism by following the previous studies on electron orbital transport~\cite{Sala2022,Choi2021, Han2022}. For parameters, we use $g \mu_B B /J = 0.25$, $E_{\rm SO} = 3$ eV, $T = 70$ K and $\partial_x T = 1$ K/$\mu$m with the constant magnon relaxation time $\tau = 30$ ns and the magnon diffusion length $\lambda = 20$ nm. The considered system size is 1$\mu$m $\times 1\mu$m. Figure~\ref{fig:4}(a) shows the resultant spin-polarized magnon orbital accumulation along the $y$-direction. We also numerically compute the electric voltage profile induced by the spin-polarized magnon orbital accumulation based on a simplified model for the inhomogeneous magnon orbital accumulation (see Supplemental Material for detailed calculation), which is shown in Fig.~\ref{fig:4}(b). Note that the estimated electric voltage at the edges is about $0.1$ $\rm \mu V$ which is within the current experimental capacity. The accumulation of the magnon orbital moment gives rise to the electric voltage profile via the magnetoelectric effect, by which we can probe the proposed magnon orbital Nernst effect.

\emph{Discussion.}\textemdash
In this Letter, we have investigated the transport of magnon orbital moments in a honeycomb AFM. The $\v k$-odd structures of both the Berry curvature and the magnon-orbital texture lead to the $\v k$-even structure of the magnon orbital Berry curvature, which gives rise to the magnon orbital Nernst effect after momentum-space integration. We emphasize that the magnon orbital Berry curvature does not require spin-orbit coupling and thus is an intrinsic property of the honeycomb AFM originating solely from the exchange interaction and the lattice geometry. As a result, the magnon orbital Nernst effect is predicted to be orders-of-magnitudes stronger than the magnon spin Nernst effect that requires spin-orbit coupling.
Although here we focus on the magnon orbital Hall effect in the honeycomb AFM,
we note that the magnon orbital Hall effect
is generally expected to be present in systems with broken inversion symmetry such as honeycomb and Kagome
ferromagnets with DMI. 
We also remark that, although we have considered a temperature gradient as a means to drive a magnon transport, one can also use electronic means to pump magnons by using, e.g., the spin Hall effects~\cite{Cornelissen2015,Goennenwein2015}.

For an experimental scheme, we have shown that the magnon orbital accumulation can be detected through the electric voltage profile by invoking the magnetoelectric effect. In particular, our theory for the magnon orbital Hall effect predicts that, upon the application of a longitudinal temperature gradient, the electric voltage profile is developed in the transverse direction. This phenomenon has the same symmetry as the Nernst effect in metallic layers. A remarkable feature of our results is that the electric voltage is not induced by the conduction electrons but by the circulating spin current associated with the orbital motion of magnons. We here note that, in addition to the electric polarization, there are several other degrees of freedom that are expected to couple with magnon orbital motions such as photons, phonons, and spin angular momenta as mentioned in Ref.~\cite{Fishman2022}, which may lead us to a new detection scheme for magnon orbitals. For the future outlook, considering a plethora of magnon-spin-related effects studied in magnonics and spintronics, it is expected that there are many magnon orbital-related effects that await the discovery involving, e.g., the interplay of the magnon orbital and the electron orbital.

\begin{acknowledgments}
We thank Kyung-Jin Lee and Giovanni Vignale for useful discussions.
G.G. acknowledges support by the National Research Foundation of Korea (NRF-2022R1C1C2006578).
H.W.L was supported by Samsung Science and Technology Foundation (SSTF-BA-1501-51).
S.K.K. was supported by Samsung Science and Technology Foundation (SSTF-BA2202-04), Brain Pool Plus Program through the National Research Foundation of Korea funded by the Ministry of Science and ICT (NRF-2020H1D3A2A03099291), and National Research Foundation of Korea funded by the Korea Government via the SRC Center for Quantum Coherence in Condensed Matter (NRF-RS-2023-00207732).
\end{acknowledgments}


\bibliography{reference}

\begin{thebibliography}{62}%
\makeatletter
\providecommand \@ifxundefined [1]{%
 \@ifx{#1\undefined}
}%
\providecommand \@ifnum [1]{%
 \ifnum #1\expandafter \@firstoftwo
 \else \expandafter \@secondoftwo
 \fi
}%
\providecommand \@ifx [1]{%
 \ifx #1\expandafter \@firstoftwo
 \else \expandafter \@secondoftwo
 \fi
}%
\providecommand \natexlab [1]{#1}%
\providecommand \enquote  [1]{``#1''}%
\providecommand \bibnamefont  [1]{#1}%
\providecommand \bibfnamefont [1]{#1}%
\providecommand \citenamefont [1]{#1}%
\providecommand \href@noop [0]{\@secondoftwo}%
\providecommand \href [0]{\begingroup \@sanitize@url \@href}%
\providecommand \@href[1]{\@@startlink{#1}\@@href}%
\providecommand \@@href[1]{\endgroup#1\@@endlink}%
\providecommand \@sanitize@url [0]{\catcode `\\12\catcode `\$12\catcode
  `\&12\catcode `\#12\catcode `\^12\catcode `\_12\catcode `\%12\relax}%
\providecommand \@@startlink[1]{}%
\providecommand \@@endlink[0]{}%
\providecommand \url  [0]{\begingroup\@sanitize@url \@url }%
\providecommand \@url [1]{\endgroup\@href {#1}{\urlprefix }}%
\providecommand \urlprefix  [0]{URL }%
\providecommand \Eprint [0]{\href }%
\providecommand \doibase [0]{https://doi.org/}%
\providecommand \selectlanguage [0]{\@gobble}%
\providecommand \bibinfo  [0]{\@secondoftwo}%
\providecommand \bibfield  [0]{\@secondoftwo}%
\providecommand \translation [1]{[#1]}%
\providecommand \BibitemOpen [0]{}%
\providecommand \bibitemStop [0]{}%
\providecommand \bibitemNoStop [0]{.\EOS\space}%
\providecommand \EOS [0]{\spacefactor3000\relax}%
\providecommand \BibitemShut  [1]{\csname bibitem#1\endcsname}%
\let\auto@bib@innerbib\@empty
\bibitem [{\citenamefont {Chumak}\ \emph {et~al.}(2015)\citenamefont {Chumak},
  \citenamefont {Vasyuchka}, \citenamefont {Serga},\ and\ \citenamefont
  {Hillebrands}}]{Chumak2015}%
  \BibitemOpen
  \bibfield  {author} {\bibinfo {author} {\bibfnamefont {A.~V.}\ \bibnamefont
  {Chumak}}, \bibinfo {author} {\bibfnamefont {V.~I.}\ \bibnamefont
  {Vasyuchka}}, \bibinfo {author} {\bibfnamefont {A.~A.}\ \bibnamefont
  {Serga}},\ and\ \bibinfo {author} {\bibfnamefont {B.}~\bibnamefont
  {Hillebrands}},\ }\bibfield  {title} {\bibinfo {title} {Magnon spintronics},\
  }\href {https://doi.org/10.1038/nphys3347} {\bibfield  {journal} {\bibinfo
  {journal} {Nat. Phys.}\ }\textbf {\bibinfo {volume} {11}},\ \bibinfo {pages}
  {453} (\bibinfo {year} {2015})}\BibitemShut {NoStop}%
\bibitem [{\citenamefont {Katsura}\ \emph {et~al.}(2010)\citenamefont
  {Katsura}, \citenamefont {Nagaosa},\ and\ \citenamefont {Lee}}]{Katsura2010}%
  \BibitemOpen
  \bibfield  {author} {\bibinfo {author} {\bibfnamefont {H.}~\bibnamefont
  {Katsura}}, \bibinfo {author} {\bibfnamefont {N.}~\bibnamefont {Nagaosa}},\
  and\ \bibinfo {author} {\bibfnamefont {P.~A.}\ \bibnamefont {Lee}},\
  }\bibfield  {title} {\bibinfo {title} {{Theory of the Thermal Hall Effect in
  Quantum Magnets}},\ }\href {https://doi.org/10.1103/PhysRevLett.104.066403}
  {\bibfield  {journal} {\bibinfo  {journal} {Phys. Rev. Lett.}\ }\textbf
  {\bibinfo {volume} {104}},\ \bibinfo {pages} {066403} (\bibinfo {year}
  {2010})}\BibitemShut {NoStop}%
\bibitem [{\citenamefont {Han}\ and\ \citenamefont {Lee}(2017)}]{Han2017}%
  \BibitemOpen
  \bibfield  {author} {\bibinfo {author} {\bibfnamefont {J.~H.}\ \bibnamefont
  {Han}}\ and\ \bibinfo {author} {\bibfnamefont {H.}~\bibnamefont {Lee}},\
  }\bibfield  {title} {\bibinfo {title} {{Spin Chirality and Hall-Like
  Transport Phenomena of Spin Excitations}},\ }\href
  {https://doi.org/10.7566/JPSJ.86.011007} {\bibfield  {journal} {\bibinfo
  {journal} {J. Phys. Soc. Jpn.}\ }\textbf {\bibinfo {volume} {86}},\ \bibinfo
  {pages} {011007} (\bibinfo {year} {2017})}\BibitemShut {NoStop}%
\bibitem [{\citenamefont {Matsumoto}\ and\ \citenamefont
  {Murakami}(2011)}]{Matsumoto2011}%
  \BibitemOpen
  \bibfield  {author} {\bibinfo {author} {\bibfnamefont {R.}~\bibnamefont
  {Matsumoto}}\ and\ \bibinfo {author} {\bibfnamefont {S.}~\bibnamefont
  {Murakami}},\ }\bibfield  {title} {\bibinfo {title} {{Theoretical Prediction
  of a Rotating Magnon Wave Packet in Ferromagnets}},\ }\href
  {https://doi.org/10.1103/PhysRevLett.106.197202} {\bibfield  {journal}
  {\bibinfo  {journal} {Phys. Rev. Lett.}\ }\textbf {\bibinfo {volume} {106}},\
  \bibinfo {pages} {197202} (\bibinfo {year} {2011})}\BibitemShut {NoStop}%
\bibitem [{\citenamefont {Onose}\ \emph {et~al.}(2010)\citenamefont {Onose},
  \citenamefont {Ideue}, \citenamefont {Katsura}, \citenamefont {Shiomi},
  \citenamefont {Nagaosa},\ and\ \citenamefont {Tokura}}]{Onose2010}%
  \BibitemOpen
  \bibfield  {author} {\bibinfo {author} {\bibfnamefont {Y.}~\bibnamefont
  {Onose}}, \bibinfo {author} {\bibfnamefont {T.}~\bibnamefont {Ideue}},
  \bibinfo {author} {\bibfnamefont {H.}~\bibnamefont {Katsura}}, \bibinfo
  {author} {\bibfnamefont {Y.}~\bibnamefont {Shiomi}}, \bibinfo {author}
  {\bibfnamefont {N.}~\bibnamefont {Nagaosa}},\ and\ \bibinfo {author}
  {\bibfnamefont {Y.}~\bibnamefont {Tokura}},\ }\bibfield  {title} {\bibinfo
  {title} {{Observation of the Magnon Hall Effect}},\ }\href
  {https://doi.org/10.1126/science.1188260} {\bibfield  {journal} {\bibinfo
  {journal} {Science}\ }\textbf {\bibinfo {volume} {329}},\ \bibinfo {pages}
  {297} (\bibinfo {year} {2010})}\BibitemShut {NoStop}%
\bibitem [{\citenamefont {Kim}\ \emph {et~al.}(2016)\citenamefont {Kim},
  \citenamefont {Ochoa}, \citenamefont {Zarzuela},\ and\ \citenamefont
  {Tserkovnyak}}]{Kim2016}%
  \BibitemOpen
  \bibfield  {author} {\bibinfo {author} {\bibfnamefont {S.~K.}\ \bibnamefont
  {Kim}}, \bibinfo {author} {\bibfnamefont {H.}~\bibnamefont {Ochoa}}, \bibinfo
  {author} {\bibfnamefont {R.}~\bibnamefont {Zarzuela}},\ and\ \bibinfo
  {author} {\bibfnamefont {Y.}~\bibnamefont {Tserkovnyak}},\ }\bibfield
  {title} {\bibinfo {title} {{Realization of the Haldane-Kane-Mele Model in a
  System of Localized Spins}},\ }\href
  {https://doi.org/10.1103/PhysRevLett.117.227201} {\bibfield  {journal}
  {\bibinfo  {journal} {Phys. Rev. Lett.}\ }\textbf {\bibinfo {volume} {117}},\
  \bibinfo {pages} {227201} (\bibinfo {year} {2016})}\BibitemShut {NoStop}%
\bibitem [{\citenamefont {Owerre}(2016)}]{Owerre2016}%
  \BibitemOpen
  \bibfield  {author} {\bibinfo {author} {\bibfnamefont {S.~A.}\ \bibnamefont
  {Owerre}},\ }\bibfield  {title} {\bibinfo {title} {{Topological honeycomb
  magnon Hall effect: A calculation of thermal Hall conductivity of magnetic
  spin excitations}},\ }\href {https://doi.org/10.1063/1.4959815} {\bibfield
  {journal} {\bibinfo  {journal} {J. Appl. Phys.}\ }\textbf {\bibinfo {volume}
  {120}},\ \bibinfo {pages} {043903} (\bibinfo {year} {2016})}\BibitemShut
  {NoStop}%
\bibitem [{\citenamefont {Cheng}\ \emph {et~al.}(2016)\citenamefont {Cheng},
  \citenamefont {Okamoto},\ and\ \citenamefont {Xiao}}]{Cheng2016}%
  \BibitemOpen
  \bibfield  {author} {\bibinfo {author} {\bibfnamefont {R.}~\bibnamefont
  {Cheng}}, \bibinfo {author} {\bibfnamefont {S.}~\bibnamefont {Okamoto}},\
  and\ \bibinfo {author} {\bibfnamefont {D.}~\bibnamefont {Xiao}},\ }\bibfield
  {title} {\bibinfo {title} {{Spin Nernst Effect of Magnons in Collinear
  Antiferromagnets}},\ }\href {https://doi.org/10.1103/PhysRevLett.117.217202}
  {\bibfield  {journal} {\bibinfo  {journal} {Phys. Rev. Lett.}\ }\textbf
  {\bibinfo {volume} {117}},\ \bibinfo {pages} {217202} (\bibinfo {year}
  {2016})}\BibitemShut {NoStop}%
\bibitem [{\citenamefont {Zyuzin}\ and\ \citenamefont
  {Kovalev}(2016)}]{Zyuzin2016}%
  \BibitemOpen
  \bibfield  {author} {\bibinfo {author} {\bibfnamefont {V.~A.}\ \bibnamefont
  {Zyuzin}}\ and\ \bibinfo {author} {\bibfnamefont {A.~A.}\ \bibnamefont
  {Kovalev}},\ }\bibfield  {title} {\bibinfo {title} {{Magnon Spin Nernst
  Effect in Antiferromagnets}},\ }\href
  {https://doi.org/10.1103/PhysRevLett.117.217203} {\bibfield  {journal}
  {\bibinfo  {journal} {Phys. Rev. Lett.}\ }\textbf {\bibinfo {volume} {117}},\
  \bibinfo {pages} {217203} (\bibinfo {year} {2016})}\BibitemShut {NoStop}%
\bibitem [{\citenamefont {Takahashi}\ and\ \citenamefont
  {Nagaosa}(2016)}]{Takahashi2016}%
  \BibitemOpen
  \bibfield  {author} {\bibinfo {author} {\bibfnamefont {R.}~\bibnamefont
  {Takahashi}}\ and\ \bibinfo {author} {\bibfnamefont {N.}~\bibnamefont
  {Nagaosa}},\ }\bibfield  {title} {\bibinfo {title} {{Berry Curvature in
  Magnon-Phonon Hybrid Systems}},\ }\href
  {https://doi.org/10.1103/PhysRevLett.117.217205} {\bibfield  {journal}
  {\bibinfo  {journal} {Phys. Rev. Lett.}\ }\textbf {\bibinfo {volume} {117}},\
  \bibinfo {pages} {217205} (\bibinfo {year} {2016})}\BibitemShut {NoStop}%
\bibitem [{\citenamefont {Zhang}\ \emph {et~al.}(2019)\citenamefont {Zhang},
  \citenamefont {Zhang}, \citenamefont {Okamoto},\ and\ \citenamefont
  {Xiao}}]{Zhang2019}%
  \BibitemOpen
  \bibfield  {author} {\bibinfo {author} {\bibfnamefont {X.}~\bibnamefont
  {Zhang}}, \bibinfo {author} {\bibfnamefont {Y.}~\bibnamefont {Zhang}},
  \bibinfo {author} {\bibfnamefont {S.}~\bibnamefont {Okamoto}},\ and\ \bibinfo
  {author} {\bibfnamefont {D.}~\bibnamefont {Xiao}},\ }\bibfield  {title}
  {\bibinfo {title} {{Thermal Hall Effect Induced by Magnon-Phonon
  Interactions}},\ }\href {https://doi.org/10.1103/PhysRevLett.123.167202}
  {\bibfield  {journal} {\bibinfo  {journal} {Phys. Rev. Lett.}\ }\textbf
  {\bibinfo {volume} {123}},\ \bibinfo {pages} {167202} (\bibinfo {year}
  {2019})}\BibitemShut {NoStop}%
\bibitem [{\citenamefont {Go}\ \emph {et~al.}(2019)\citenamefont {Go},
  \citenamefont {Kim},\ and\ \citenamefont {Lee}}]{Go2019}%
  \BibitemOpen
  \bibfield  {author} {\bibinfo {author} {\bibfnamefont {G.}~\bibnamefont
  {Go}}, \bibinfo {author} {\bibfnamefont {S.~K.}\ \bibnamefont {Kim}},\ and\
  \bibinfo {author} {\bibfnamefont {K.-J.}\ \bibnamefont {Lee}},\ }\bibfield
  {title} {\bibinfo {title} {{Topological Magnon-Phonon Hybrid Excitations in
  Two-Dimensional Ferromagnets with Tunable Chern Numbers}},\ }\href
  {https://doi.org/10.1103/PhysRevLett.123.237207} {\bibfield  {journal}
  {\bibinfo  {journal} {Phys. Rev. Lett.}\ }\textbf {\bibinfo {volume} {123}},\
  \bibinfo {pages} {237207} (\bibinfo {year} {2019})}\BibitemShut {NoStop}%
\bibitem [{\citenamefont {Park}\ \emph {et~al.}(2020)\citenamefont {Park},
  \citenamefont {Nagaosa},\ and\ \citenamefont {Yang}}]{Park2020}%
  \BibitemOpen
  \bibfield  {author} {\bibinfo {author} {\bibfnamefont {S.}~\bibnamefont
  {Park}}, \bibinfo {author} {\bibfnamefont {N.}~\bibnamefont {Nagaosa}},\ and\
  \bibinfo {author} {\bibfnamefont {B.-J.}\ \bibnamefont {Yang}},\ }\bibfield
  {title} {\bibinfo {title} {{Thermal Hall Effect, Spin Nernst Effect, and Spin
  Density Induced by a Thermal Gradient in Collinear Ferrimagnets from
  Magnon--Phonon Interaction}},\ }\href
  {https://doi.org/10.1021/acs.nanolett.0c00363} {\bibfield  {journal}
  {\bibinfo  {journal} {Nano Lett.}\ }\textbf {\bibinfo {volume} {20}},\
  \bibinfo {pages} {2741} (\bibinfo {year} {2020})}\BibitemShut {NoStop}%
\bibitem [{\citenamefont {Tanaka}\ \emph {et~al.}(2008)\citenamefont {Tanaka},
  \citenamefont {Kontani}, \citenamefont {Naito}, \citenamefont {Naito},
  \citenamefont {Hirashima}, \citenamefont {Yamada},\ and\ \citenamefont
  {Inoue}}]{Tanaka2008}%
  \BibitemOpen
  \bibfield  {author} {\bibinfo {author} {\bibfnamefont {T.}~\bibnamefont
  {Tanaka}}, \bibinfo {author} {\bibfnamefont {H.}~\bibnamefont {Kontani}},
  \bibinfo {author} {\bibfnamefont {M.}~\bibnamefont {Naito}}, \bibinfo
  {author} {\bibfnamefont {T.}~\bibnamefont {Naito}}, \bibinfo {author}
  {\bibfnamefont {D.~S.}\ \bibnamefont {Hirashima}}, \bibinfo {author}
  {\bibfnamefont {K.}~\bibnamefont {Yamada}},\ and\ \bibinfo {author}
  {\bibfnamefont {J.}~\bibnamefont {Inoue}},\ }\bibfield  {title} {\bibinfo
  {title} {{Intrinsic spin Hall effect and orbital Hall effect in $4d$ and $5d$
  transition metals}},\ }\href {https://doi.org/10.1103/PhysRevB.77.165117}
  {\bibfield  {journal} {\bibinfo  {journal} {Phys. Rev. B}\ }\textbf {\bibinfo
  {volume} {77}},\ \bibinfo {pages} {165117} (\bibinfo {year}
  {2008})}\BibitemShut {NoStop}%
\bibitem [{\citenamefont {Kontani}\ \emph {et~al.}(2008)\citenamefont
  {Kontani}, \citenamefont {Tanaka}, \citenamefont {Hirashima}, \citenamefont
  {Yamada},\ and\ \citenamefont {Inoue}}]{Kontani2008}%
  \BibitemOpen
  \bibfield  {author} {\bibinfo {author} {\bibfnamefont {H.}~\bibnamefont
  {Kontani}}, \bibinfo {author} {\bibfnamefont {T.}~\bibnamefont {Tanaka}},
  \bibinfo {author} {\bibfnamefont {D.~S.}\ \bibnamefont {Hirashima}}, \bibinfo
  {author} {\bibfnamefont {K.}~\bibnamefont {Yamada}},\ and\ \bibinfo {author}
  {\bibfnamefont {J.}~\bibnamefont {Inoue}},\ }\bibfield  {title} {\bibinfo
  {title} {{Giant Intrinsic Spin and Orbital Hall Effects in
  ${\mathrm{Sr}}_{2}M{\mathrm{O}}_{4}$ ($M=\mathrm{Ru}$, Rh, Mo)}},\ }\href
  {https://doi.org/10.1103/PhysRevLett.100.096601} {\bibfield  {journal}
  {\bibinfo  {journal} {Phys. Rev. Lett.}\ }\textbf {\bibinfo {volume} {100}},\
  \bibinfo {pages} {096601} (\bibinfo {year} {2008})}\BibitemShut {NoStop}%
\bibitem [{\citenamefont {Kontani}\ \emph {et~al.}(2009)\citenamefont
  {Kontani}, \citenamefont {Tanaka}, \citenamefont {Hirashima}, \citenamefont
  {Yamada},\ and\ \citenamefont {Inoue}}]{Kontani2009}%
  \BibitemOpen
  \bibfield  {author} {\bibinfo {author} {\bibfnamefont {H.}~\bibnamefont
  {Kontani}}, \bibinfo {author} {\bibfnamefont {T.}~\bibnamefont {Tanaka}},
  \bibinfo {author} {\bibfnamefont {D.~S.}\ \bibnamefont {Hirashima}}, \bibinfo
  {author} {\bibfnamefont {K.}~\bibnamefont {Yamada}},\ and\ \bibinfo {author}
  {\bibfnamefont {J.}~\bibnamefont {Inoue}},\ }\bibfield  {title} {\bibinfo
  {title} {{Giant Orbital Hall Effect in Transition Metals: Origin of Large
  Spin and Anomalous Hall Effects}},\ }\href
  {https://doi.org/10.1103/PhysRevLett.102.016601} {\bibfield  {journal}
  {\bibinfo  {journal} {Phys. Rev. Lett.}\ }\textbf {\bibinfo {volume} {102}},\
  \bibinfo {pages} {016601} (\bibinfo {year} {2009})}\BibitemShut {NoStop}%
\bibitem [{\citenamefont {Tanaka}\ and\ \citenamefont
  {Kontani}(2010)}]{Tanaka2010}%
  \BibitemOpen
  \bibfield  {author} {\bibinfo {author} {\bibfnamefont {T.}~\bibnamefont
  {Tanaka}}\ and\ \bibinfo {author} {\bibfnamefont {H.}~\bibnamefont
  {Kontani}},\ }\bibfield  {title} {\bibinfo {title} {{Intrinsic spin and
  orbital Hall effects in heavy-fermion systems}},\ }\href
  {https://doi.org/10.1103/PhysRevB.81.224401} {\bibfield  {journal} {\bibinfo
  {journal} {Phys. Rev. B}\ }\textbf {\bibinfo {volume} {81}},\ \bibinfo
  {pages} {224401} (\bibinfo {year} {2010})}\BibitemShut {NoStop}%
\bibitem [{\citenamefont {Go}\ \emph {et~al.}(2018)\citenamefont {Go},
  \citenamefont {Jo}, \citenamefont {Kim},\ and\ \citenamefont {Lee}}]{Go2018}%
  \BibitemOpen
  \bibfield  {author} {\bibinfo {author} {\bibfnamefont {D.}~\bibnamefont
  {Go}}, \bibinfo {author} {\bibfnamefont {D.}~\bibnamefont {Jo}}, \bibinfo
  {author} {\bibfnamefont {C.}~\bibnamefont {Kim}},\ and\ \bibinfo {author}
  {\bibfnamefont {H.-W.}\ \bibnamefont {Lee}},\ }\bibfield  {title} {\bibinfo
  {title} {{Intrinsic Spin and Orbital Hall Effects from Orbital Texture}},\
  }\href {https://doi.org/10.1103/PhysRevLett.121.086602} {\bibfield  {journal}
  {\bibinfo  {journal} {Phys. Rev. Lett.}\ }\textbf {\bibinfo {volume} {121}},\
  \bibinfo {pages} {086602} (\bibinfo {year} {2018})}\BibitemShut {NoStop}%
\bibitem [{\citenamefont {Jo}\ \emph {et~al.}(2018)\citenamefont {Jo},
  \citenamefont {Go},\ and\ \citenamefont {Lee}}]{Jo2018}%
  \BibitemOpen
  \bibfield  {author} {\bibinfo {author} {\bibfnamefont {D.}~\bibnamefont
  {Jo}}, \bibinfo {author} {\bibfnamefont {D.}~\bibnamefont {Go}},\ and\
  \bibinfo {author} {\bibfnamefont {H.-W.}\ \bibnamefont {Lee}},\ }\bibfield
  {title} {\bibinfo {title} {{Gigantic intrinsic orbital Hall effects in weakly
  spin-orbit coupled metals}},\ }\href
  {https://doi.org/10.1103/PhysRevB.98.214405} {\bibfield  {journal} {\bibinfo
  {journal} {Phys. Rev. B}\ }\textbf {\bibinfo {volume} {98}},\ \bibinfo
  {pages} {214405} (\bibinfo {year} {2018})}\BibitemShut {NoStop}%
\bibitem [{\citenamefont {Sala}\ and\ \citenamefont
  {Gambardella}(2022)}]{Sala2022}%
  \BibitemOpen
  \bibfield  {author} {\bibinfo {author} {\bibfnamefont {G.}~\bibnamefont
  {Sala}}\ and\ \bibinfo {author} {\bibfnamefont {P.}~\bibnamefont
  {Gambardella}},\ }\bibfield  {title} {\bibinfo {title} {{Giant orbital Hall
  effect and orbital-to-spin conversion in $3d$, $5d$, and $4f$ metallic
  heterostructures}},\ }\href
  {https://doi.org/10.1103/PhysRevResearch.4.033037} {\bibfield  {journal}
  {\bibinfo  {journal} {Phys. Rev. Res.}\ }\textbf {\bibinfo {volume} {4}},\
  \bibinfo {pages} {033037} (\bibinfo {year} {2022})}\BibitemShut {NoStop}%
\bibitem [{\citenamefont {Bhowal}\ and\ \citenamefont
  {Vignale}(2021)}]{Bhowal2021}%
  \BibitemOpen
  \bibfield  {author} {\bibinfo {author} {\bibfnamefont {S.}~\bibnamefont
  {Bhowal}}\ and\ \bibinfo {author} {\bibfnamefont {G.}~\bibnamefont
  {Vignale}},\ }\bibfield  {title} {\bibinfo {title} {{Orbital Hall effect as
  an alternative to valley Hall effect in gapped graphene}},\ }\href
  {https://doi.org/10.1103/PhysRevB.103.195309} {\bibfield  {journal} {\bibinfo
   {journal} {Phys. Rev. B}\ }\textbf {\bibinfo {volume} {103}},\ \bibinfo
  {pages} {195309} (\bibinfo {year} {2021})}\BibitemShut {NoStop}%
\bibitem [{\citenamefont {Pezo}\ \emph {et~al.}(2022)\citenamefont {Pezo},
  \citenamefont {Garc\'{\i}a~Ovalle},\ and\ \citenamefont
  {Manchon}}]{Pezo2022}%
  \BibitemOpen
  \bibfield  {author} {\bibinfo {author} {\bibfnamefont {A.}~\bibnamefont
  {Pezo}}, \bibinfo {author} {\bibfnamefont {D.}~\bibnamefont
  {Garc\'{\i}a~Ovalle}},\ and\ \bibinfo {author} {\bibfnamefont
  {A.}~\bibnamefont {Manchon}},\ }\bibfield  {title} {\bibinfo {title}
  {{Orbital Hall effect in crystals: Interatomic versus intra-atomic
  contributions}},\ }\href {https://doi.org/10.1103/PhysRevB.106.104414}
  {\bibfield  {journal} {\bibinfo  {journal} {Phys. Rev. B}\ }\textbf {\bibinfo
  {volume} {106}},\ \bibinfo {pages} {104414} (\bibinfo {year}
  {2022})}\BibitemShut {NoStop}%
\bibitem [{\citenamefont {Go}\ and\ \citenamefont {Lee}(2020)}]{Go2020PRR}%
  \BibitemOpen
  \bibfield  {author} {\bibinfo {author} {\bibfnamefont {D.}~\bibnamefont
  {Go}}\ and\ \bibinfo {author} {\bibfnamefont {H.-W.}\ \bibnamefont {Lee}},\
  }\bibfield  {title} {\bibinfo {title} {{Orbital torque: Torque generation by
  orbital current injection}},\ }\href
  {https://doi.org/10.1103/PhysRevResearch.2.013177} {\bibfield  {journal}
  {\bibinfo  {journal} {Phys. Rev. Res.}\ }\textbf {\bibinfo {volume} {2}},\
  \bibinfo {pages} {013177} (\bibinfo {year} {2020})}\BibitemShut {NoStop}%
\bibitem [{\citenamefont {Lee}\ \emph {et~al.}(2021)\citenamefont {Lee},
  \citenamefont {Go}, \citenamefont {Park}, \citenamefont {Jeong},
  \citenamefont {Ko}, \citenamefont {Yun}, \citenamefont {Jo}, \citenamefont
  {Lee}, \citenamefont {Go}, \citenamefont {Oh}, \citenamefont {Kim},
  \citenamefont {Park}, \citenamefont {Min}, \citenamefont {Koo}, \citenamefont
  {Lee}, \citenamefont {Lee},\ and\ \citenamefont {Lee}}]{Lee2021}%
  \BibitemOpen
  \bibfield  {author} {\bibinfo {author} {\bibfnamefont {D.}~\bibnamefont
  {Lee}}, \bibinfo {author} {\bibfnamefont {D.}~\bibnamefont {Go}}, \bibinfo
  {author} {\bibfnamefont {H.-J.}\ \bibnamefont {Park}}, \bibinfo {author}
  {\bibfnamefont {W.}~\bibnamefont {Jeong}}, \bibinfo {author} {\bibfnamefont
  {H.-W.}\ \bibnamefont {Ko}}, \bibinfo {author} {\bibfnamefont
  {D.}~\bibnamefont {Yun}}, \bibinfo {author} {\bibfnamefont {D.}~\bibnamefont
  {Jo}}, \bibinfo {author} {\bibfnamefont {S.}~\bibnamefont {Lee}}, \bibinfo
  {author} {\bibfnamefont {G.}~\bibnamefont {Go}}, \bibinfo {author}
  {\bibfnamefont {J.~H.}\ \bibnamefont {Oh}}, \bibinfo {author} {\bibfnamefont
  {K.-J.}\ \bibnamefont {Kim}}, \bibinfo {author} {\bibfnamefont {B.-G.}\
  \bibnamefont {Park}}, \bibinfo {author} {\bibfnamefont {B.-C.}\ \bibnamefont
  {Min}}, \bibinfo {author} {\bibfnamefont {H.~C.}\ \bibnamefont {Koo}},
  \bibinfo {author} {\bibfnamefont {H.-W.}\ \bibnamefont {Lee}}, \bibinfo
  {author} {\bibfnamefont {O.}~\bibnamefont {Lee}},\ and\ \bibinfo {author}
  {\bibfnamefont {K.-J.}\ \bibnamefont {Lee}},\ }\bibfield  {title} {\bibinfo
  {title} {{Orbital torque in magnetic bilayers}},\ }\href
  {https://doi.org/https://doi.org/10.1038/s41467-021-26650-9} {\bibfield
  {journal} {\bibinfo  {journal} {Nat. Commun.}\ }\textbf {\bibinfo {volume}
  {12}},\ \bibinfo {pages} {6710} (\bibinfo {year} {2021})}\BibitemShut
  {NoStop}%
\bibitem [{\citenamefont {Hayashi}\ \emph {et~al.}()\citenamefont {Hayashi},
  \citenamefont {Jo}, \citenamefont {Go}, \citenamefont {Mokrousov},
  \citenamefont {Lee},\ and\ \citenamefont {Ando}}]{Hayashi2022}%
  \BibitemOpen
  \bibfield  {author} {\bibinfo {author} {\bibfnamefont {H.}~\bibnamefont
  {Hayashi}}, \bibinfo {author} {\bibfnamefont {D.}~\bibnamefont {Jo}},
  \bibinfo {author} {\bibfnamefont {D.}~\bibnamefont {Go}}, \bibinfo {author}
  {\bibfnamefont {Y.}~\bibnamefont {Mokrousov}}, \bibinfo {author}
  {\bibfnamefont {H.-W.}\ \bibnamefont {Lee}},\ and\ \bibinfo {author}
  {\bibfnamefont {K.}~\bibnamefont {Ando}},\ }\href@noop {} {\bibinfo {title}
  {{Observation of long-range orbital transport and giant orbital torque}}},\
  \Eprint {https://arxiv.org/abs/arXiv:2202.13896} {arXiv:2202.13896}
  \BibitemShut {NoStop}%
\bibitem [{\citenamefont {Haigh}\ \emph {et~al.}(2016)\citenamefont {Haigh},
  \citenamefont {Nunnenkamp}, \citenamefont {Ramsay},\ and\ \citenamefont
  {Ferguson}}]{Haigh2016}%
  \BibitemOpen
  \bibfield  {author} {\bibinfo {author} {\bibfnamefont {J.~A.}\ \bibnamefont
  {Haigh}}, \bibinfo {author} {\bibfnamefont {A.}~\bibnamefont {Nunnenkamp}},
  \bibinfo {author} {\bibfnamefont {A.~J.}\ \bibnamefont {Ramsay}},\ and\
  \bibinfo {author} {\bibfnamefont {A.~J.}\ \bibnamefont {Ferguson}},\
  }\bibfield  {title} {\bibinfo {title} {{Triple-Resonant Brillouin Light
  Scattering in Magneto-Optical Cavities}},\ }\href
  {https://doi.org/10.1103/PhysRevLett.117.133602} {\bibfield  {journal}
  {\bibinfo  {journal} {Phys. Rev. Lett.}\ }\textbf {\bibinfo {volume} {117}},\
  \bibinfo {pages} {133602} (\bibinfo {year} {2016})}\BibitemShut {NoStop}%
\bibitem [{\citenamefont {Sharma}\ \emph {et~al.}(2017)\citenamefont {Sharma},
  \citenamefont {Blanter},\ and\ \citenamefont {Bauer}}]{Sharma2017}%
  \BibitemOpen
  \bibfield  {author} {\bibinfo {author} {\bibfnamefont {S.}~\bibnamefont
  {Sharma}}, \bibinfo {author} {\bibfnamefont {Y.~M.}\ \bibnamefont
  {Blanter}},\ and\ \bibinfo {author} {\bibfnamefont {G.~E.~W.}\ \bibnamefont
  {Bauer}},\ }\bibfield  {title} {\bibinfo {title} {{Light scattering by
  magnons in whispering gallery mode cavities}},\ }\href
  {https://doi.org/10.1103/PhysRevB.96.094412} {\bibfield  {journal} {\bibinfo
  {journal} {Phys. Rev. B}\ }\textbf {\bibinfo {volume} {96}},\ \bibinfo
  {pages} {094412} (\bibinfo {year} {2017})}\BibitemShut {NoStop}%
\bibitem [{\citenamefont {Osada}\ \emph {et~al.}(2018)\citenamefont {Osada},
  \citenamefont {Gloppe}, \citenamefont {Nakamura},\ and\ \citenamefont
  {Usami}}]{Osada2018}%
  \BibitemOpen
  \bibfield  {author} {\bibinfo {author} {\bibfnamefont {A.}~\bibnamefont
  {Osada}}, \bibinfo {author} {\bibfnamefont {A.}~\bibnamefont {Gloppe}},
  \bibinfo {author} {\bibfnamefont {Y.}~\bibnamefont {Nakamura}},\ and\
  \bibinfo {author} {\bibfnamefont {K.}~\bibnamefont {Usami}},\ }\bibfield
  {title} {\bibinfo {title} {{Orbital angular momentum conservation in
  Brillouin light scattering within a ferromagnetic sphere}},\ }\href
  {https://doi.org/10.1088/1367-2630/aae4b1} {\bibfield  {journal} {\bibinfo
  {journal} {New J. Phys.}\ }\textbf {\bibinfo {volume} {20}},\ \bibinfo
  {pages} {103018} (\bibinfo {year} {2018})}\BibitemShut {NoStop}%
\bibitem [{\citenamefont {Jiang}\ \emph {et~al.}(2020)\citenamefont {Jiang},
  \citenamefont {Yuan}, \citenamefont {Li}, \citenamefont {Wang}, \citenamefont
  {Zhang}, \citenamefont {Cao},\ and\ \citenamefont {Yan}}]{Jiang2020}%
  \BibitemOpen
  \bibfield  {author} {\bibinfo {author} {\bibfnamefont {Y.}~\bibnamefont
  {Jiang}}, \bibinfo {author} {\bibfnamefont {H.~Y.}\ \bibnamefont {Yuan}},
  \bibinfo {author} {\bibfnamefont {Z.-X.}\ \bibnamefont {Li}}, \bibinfo
  {author} {\bibfnamefont {Z.}~\bibnamefont {Wang}}, \bibinfo {author}
  {\bibfnamefont {H.~W.}\ \bibnamefont {Zhang}}, \bibinfo {author}
  {\bibfnamefont {Y.}~\bibnamefont {Cao}},\ and\ \bibinfo {author}
  {\bibfnamefont {P.}~\bibnamefont {Yan}},\ }\bibfield  {title} {\bibinfo
  {title} {{Twisted Magnon as a Magnetic Tweezer}},\ }\href
  {https://doi.org/10.1103/PhysRevLett.124.217204} {\bibfield  {journal}
  {\bibinfo  {journal} {Phys. Rev. Lett.}\ }\textbf {\bibinfo {volume} {124}},\
  \bibinfo {pages} {217204} (\bibinfo {year} {2020})}\BibitemShut {NoStop}%
\bibitem [{\citenamefont {Gonz{\'a}lez}\ \emph {et~al.}(2010)\citenamefont
  {Gonz{\'a}lez}, \citenamefont {Landeros},\ and\ \citenamefont
  {N{\'u}{\~n}ez}}]{GonzalezJMMM2010}%
  \BibitemOpen
  \bibfield  {author} {\bibinfo {author} {\bibfnamefont {A.}~\bibnamefont
  {Gonz{\'a}lez}}, \bibinfo {author} {\bibfnamefont {P.}~\bibnamefont
  {Landeros}},\ and\ \bibinfo {author} {\bibfnamefont {{\'A}.~S.}\ \bibnamefont
  {N{\'u}{\~n}ez}},\ }\bibfield  {title} {\bibinfo {title} {Spin wave spectrum
  of magnetic nanotubes},\ }\href
  {https://doi.org/https://doi.org/10.1016/j.jmmm.2009.10.010} {\bibfield
  {journal} {\bibinfo  {journal} {J. Magn. Magn. Mater.}\ }\textbf {\bibinfo
  {volume} {322}},\ \bibinfo {pages} {530 } (\bibinfo {year}
  {2010})}\BibitemShut {NoStop}%
\bibitem [{\citenamefont {Lee}\ and\ \citenamefont {Kim}(2021)}]{LeeSH2021}%
  \BibitemOpen
  \bibfield  {author} {\bibinfo {author} {\bibfnamefont {S.}~\bibnamefont
  {Lee}}\ and\ \bibinfo {author} {\bibfnamefont {S.~K.}\ \bibnamefont {Kim}},\
  }\bibfield  {title} {\bibinfo {title} {{Orbital angular momentum and
  current-induced motion of a topologically textured domain wall in a
  ferromagnetic nanotube}},\ }\href
  {https://doi.org/10.1103/PhysRevB.104.L140401} {\bibfield  {journal}
  {\bibinfo  {journal} {Phys. Rev. B}\ }\textbf {\bibinfo {volume} {104}},\
  \bibinfo {pages} {L140401} (\bibinfo {year} {2021})}\BibitemShut {NoStop}%
\bibitem [{\citenamefont {Lee}\ and\ \citenamefont {Kim}(2022)}]{LeeSH2022}%
  \BibitemOpen
  \bibfield  {author} {\bibinfo {author} {\bibfnamefont {S.}~\bibnamefont
  {Lee}}\ and\ \bibinfo {author} {\bibfnamefont {S.~K.}\ \bibnamefont {Kim}},\
  }\bibfield  {title} {\bibinfo {title} {{Generation of Magnon Orbital Angular
  Momentum by a Skyrmion-Textured Domain Wall in a Ferromagnetic Nanotube}},\
  }\href {https://doi.org/10.3389/fphy.2022.858614} {\bibfield  {journal}
  {\bibinfo  {journal} {Front. Phys.}\ }\textbf {\bibinfo {volume} {10}},\
  \bibinfo {pages} {858614} (\bibinfo {year} {2022})}\BibitemShut {NoStop}%
\bibitem [{\citenamefont {Neumann}\ \emph {et~al.}(2020)\citenamefont
  {Neumann}, \citenamefont {Mook}, \citenamefont {Henk},\ and\ \citenamefont
  {Mertig}}]{Neumann2000}%
  \BibitemOpen
  \bibfield  {author} {\bibinfo {author} {\bibfnamefont {R.~R.}\ \bibnamefont
  {Neumann}}, \bibinfo {author} {\bibfnamefont {A.}~\bibnamefont {Mook}},
  \bibinfo {author} {\bibfnamefont {J.}~\bibnamefont {Henk}},\ and\ \bibinfo
  {author} {\bibfnamefont {I.}~\bibnamefont {Mertig}},\ }\bibfield  {title}
  {\bibinfo {title} {{Orbital Magnetic Moment of Magnons}},\ }\href
  {https://doi.org/10.1103/PhysRevLett.125.117209} {\bibfield  {journal}
  {\bibinfo  {journal} {Phys. Rev. Lett.}\ }\textbf {\bibinfo {volume} {125}},\
  \bibinfo {pages} {117209} (\bibinfo {year} {2020})}\BibitemShut {NoStop}%
\bibitem [{\citenamefont {Fishman}\ \emph
  {et~al.}(2022{\natexlab{a}})\citenamefont {Fishman}, \citenamefont
  {Gardner},\ and\ \citenamefont {Okamoto}}]{Fishman2022}%
  \BibitemOpen
  \bibfield  {author} {\bibinfo {author} {\bibfnamefont {R.~S.}\ \bibnamefont
  {Fishman}}, \bibinfo {author} {\bibfnamefont {J.~S.}\ \bibnamefont
  {Gardner}},\ and\ \bibinfo {author} {\bibfnamefont {S.}~\bibnamefont
  {Okamoto}},\ }\bibfield  {title} {\bibinfo {title} {{Orbital Angular Momentum
  of Magnons in Collinear Magnets}},\ }\href
  {https://doi.org/10.1103/PhysRevLett.129.167202} {\bibfield  {journal}
  {\bibinfo  {journal} {Phys. Rev. Lett.}\ }\textbf {\bibinfo {volume} {129}},\
  \bibinfo {pages} {167202} (\bibinfo {year} {2022}{\natexlab{a}})}\BibitemShut
  {NoStop}%
\bibitem [{\citenamefont {Fishman}\ \emph
  {et~al.}(2022{\natexlab{b}})\citenamefont {Fishman}, \citenamefont
  {Lindsay},\ and\ \citenamefont {Okamoto}}]{Fishman2022b}%
  \BibitemOpen
  \bibfield  {author} {\bibinfo {author} {\bibfnamefont {R.~S.}\ \bibnamefont
  {Fishman}}, \bibinfo {author} {\bibfnamefont {L.}~\bibnamefont {Lindsay}},\
  and\ \bibinfo {author} {\bibfnamefont {S.}~\bibnamefont {Okamoto}},\
  }\bibfield  {title} {\bibinfo {title} {{Exact results for the orbital angular
  momentum of magnons on honeycomb lattices}},\ }\href
  {https://doi.org/10.1088/1361-648X/ac9a28} {\bibfield  {journal} {\bibinfo
  {journal} {J. Phys. Condens. Matter}\ }\textbf {\bibinfo {volume} {35}},\
  \bibinfo {pages} {015801} (\bibinfo {year} {2022}{\natexlab{b}})}\BibitemShut
  {NoStop}%
\bibitem [{\citenamefont {Wu}\ and\ \citenamefont {Hu}(2015)}]{Wu2015}%
  \BibitemOpen
  \bibfield  {author} {\bibinfo {author} {\bibfnamefont {L.-H.}\ \bibnamefont
  {Wu}}\ and\ \bibinfo {author} {\bibfnamefont {X.}~\bibnamefont {Hu}},\
  }\bibfield  {title} {\bibinfo {title} {{Scheme for Achieving a Topological
  Photonic Crystal by Using Dielectric Material}},\ }\href
  {https://doi.org/10.1103/PhysRevLett.114.223901} {\bibfield  {journal}
  {\bibinfo  {journal} {Phys. Rev. Lett.}\ }\textbf {\bibinfo {volume} {114}},\
  \bibinfo {pages} {223901} (\bibinfo {year} {2015})}\BibitemShut {NoStop}%
\bibitem [{\citenamefont {He}\ \emph {et~al.}(2016)\citenamefont {He},
  \citenamefont {Ni}, \citenamefont {Ge}, \citenamefont {Sun}, \citenamefont
  {Chen}, \citenamefont {Lu}, \citenamefont {Liu},\ and\ \citenamefont
  {Chen}}]{He2016}%
  \BibitemOpen
  \bibfield  {author} {\bibinfo {author} {\bibfnamefont {C.}~\bibnamefont
  {He}}, \bibinfo {author} {\bibfnamefont {X.}~\bibnamefont {Ni}}, \bibinfo
  {author} {\bibfnamefont {H.}~\bibnamefont {Ge}}, \bibinfo {author}
  {\bibfnamefont {X.-C.}\ \bibnamefont {Sun}}, \bibinfo {author} {\bibfnamefont
  {Y.-B.}\ \bibnamefont {Chen}}, \bibinfo {author} {\bibfnamefont {M.-H.}\
  \bibnamefont {Lu}}, \bibinfo {author} {\bibfnamefont {X.-P.}\ \bibnamefont
  {Liu}},\ and\ \bibinfo {author} {\bibfnamefont {Y.-F.}\ \bibnamefont
  {Chen}},\ }\bibfield  {title} {\bibinfo {title} {{Acoustic topological
  insulator and robust one-way sound transport}},\ }\href
  {https://doi.org/https://doi.org/10.1038/nphys3867} {\bibfield  {journal}
  {\bibinfo  {journal} {Nat. Phys.}\ }\textbf {\bibinfo {volume} {12}},\
  \bibinfo {pages} {1124} (\bibinfo {year} {2016})}\BibitemShut {NoStop}%
\bibitem [{\citenamefont {Li}\ \emph {et~al.}(2018)\citenamefont {Li},
  \citenamefont {Sun}, \citenamefont {Zhu}, \citenamefont {Guo}, \citenamefont
  {Jiang}, \citenamefont {Kariyado}, \citenamefont {Chen},\ and\ \citenamefont
  {Hu}}]{Li2018}%
  \BibitemOpen
  \bibfield  {author} {\bibinfo {author} {\bibfnamefont {Y.}~\bibnamefont
  {Li}}, \bibinfo {author} {\bibfnamefont {Y.}~\bibnamefont {Sun}}, \bibinfo
  {author} {\bibfnamefont {W.}~\bibnamefont {Zhu}}, \bibinfo {author}
  {\bibfnamefont {Z.}~\bibnamefont {Guo}}, \bibinfo {author} {\bibfnamefont
  {J.}~\bibnamefont {Jiang}}, \bibinfo {author} {\bibfnamefont
  {T.}~\bibnamefont {Kariyado}}, \bibinfo {author} {\bibfnamefont
  {H.}~\bibnamefont {Chen}},\ and\ \bibinfo {author} {\bibfnamefont
  {X.}~\bibnamefont {Hu}},\ }\bibfield  {title} {\bibinfo {title} {{Topological
  {LC-circuits} based on microstrips and observation of electromagnetic modes
  with orbital angular momentum}},\ }\href
  {https://doi.org/https://doi.org/10.1038/s41467-018-07084-2} {\bibfield
  {journal} {\bibinfo  {journal} {Nat. Commun.}\ }\textbf {\bibinfo {volume}
  {9}},\ \bibinfo {pages} {4598} (\bibinfo {year} {2018})}\BibitemShut
  {NoStop}%
\bibitem [{\citenamefont {Huang}\ \emph {et~al.}(2022)\citenamefont {Huang},
  \citenamefont {Kariyado},\ and\ \citenamefont {Hu}}]{Huang2022}%
  \BibitemOpen
  \bibfield  {author} {\bibinfo {author} {\bibfnamefont {H.}~\bibnamefont
  {Huang}}, \bibinfo {author} {\bibfnamefont {T.}~\bibnamefont {Kariyado}},\
  and\ \bibinfo {author} {\bibfnamefont {X.}~\bibnamefont {Hu}},\ }\bibfield
  {title} {\bibinfo {title} {{Topological magnon modes on honeycomb lattice
  with coupling textures}},\ }\href {https://doi.org/Topological magnon modes
  on honeycomb lattice with coupling textures} {\bibfield  {journal} {\bibinfo
  {journal} {Sci. Rep.}\ }\textbf {\bibinfo {volume} {12}},\ \bibinfo {pages}
  {6257} (\bibinfo {year} {2022})}\BibitemShut {NoStop}%
\bibitem [{\citenamefont {Katsura}\ \emph {et~al.}(2005)\citenamefont
  {Katsura}, \citenamefont {Nagaosa},\ and\ \citenamefont
  {Balatsky}}]{Katsura2005}%
  \BibitemOpen
  \bibfield  {author} {\bibinfo {author} {\bibfnamefont {H.}~\bibnamefont
  {Katsura}}, \bibinfo {author} {\bibfnamefont {N.}~\bibnamefont {Nagaosa}},\
  and\ \bibinfo {author} {\bibfnamefont {A.~V.}\ \bibnamefont {Balatsky}},\
  }\bibfield  {title} {\bibinfo {title} {{Spin Current and Magnetoelectric
  Effect in Noncollinear Magnets}},\ }\href
  {https://doi.org/10.1103/PhysRevLett.95.057205} {\bibfield  {journal}
  {\bibinfo  {journal} {Phys. Rev. Lett.}\ }\textbf {\bibinfo {volume} {95}},\
  \bibinfo {pages} {057205} (\bibinfo {year} {2005})}\BibitemShut {NoStop}%
\bibitem [{\citenamefont {Liu}\ and\ \citenamefont
  {Vignale}(2011)}]{Vignale2011}%
  \BibitemOpen
  \bibfield  {author} {\bibinfo {author} {\bibfnamefont {T.}~\bibnamefont
  {Liu}}\ and\ \bibinfo {author} {\bibfnamefont {G.}~\bibnamefont {Vignale}},\
  }\bibfield  {title} {\bibinfo {title} {{Electric Control of Spin Currents and
  Spin-Wave Logic}},\ }\href {https://doi.org/10.1103/PhysRevLett.106.247203}
  {\bibfield  {journal} {\bibinfo  {journal} {Phys. Rev. Lett.}\ }\textbf
  {\bibinfo {volume} {106}},\ \bibinfo {pages} {247203} (\bibinfo {year}
  {2011})}\BibitemShut {NoStop}%
\bibitem [{\citenamefont {Wang}\ \emph {et~al.}(2016)\citenamefont {Wang},
  \citenamefont {Du}, \citenamefont {Liu}, \citenamefont {Hu}, \citenamefont
  {Zhang}, \citenamefont {Zhang}, \citenamefont {Owen}, \citenamefont {Lu},
  \citenamefont {Gan}, \citenamefont {Sengupta}, \citenamefont {Kloc},\ and\
  \citenamefont {Xiong}}]{Wang2016}%
  \BibitemOpen
  \bibfield  {author} {\bibinfo {author} {\bibfnamefont {X.}~\bibnamefont
  {Wang}}, \bibinfo {author} {\bibfnamefont {K.}~\bibnamefont {Du}}, \bibinfo
  {author} {\bibfnamefont {Y.~Y.~F.}\ \bibnamefont {Liu}}, \bibinfo {author}
  {\bibfnamefont {P.}~\bibnamefont {Hu}}, \bibinfo {author} {\bibfnamefont
  {J.}~\bibnamefont {Zhang}}, \bibinfo {author} {\bibfnamefont
  {Q.}~\bibnamefont {Zhang}}, \bibinfo {author} {\bibfnamefont {M.~H.~S.}\
  \bibnamefont {Owen}}, \bibinfo {author} {\bibfnamefont {X.}~\bibnamefont
  {Lu}}, \bibinfo {author} {\bibfnamefont {C.~K.}\ \bibnamefont {Gan}},
  \bibinfo {author} {\bibfnamefont {P.}~\bibnamefont {Sengupta}}, \bibinfo
  {author} {\bibfnamefont {C.}~\bibnamefont {Kloc}},\ and\ \bibinfo {author}
  {\bibfnamefont {Q.}~\bibnamefont {Xiong}},\ }\bibfield  {title} {\bibinfo
  {title} {{Raman spectroscopy of atomically thin two-dimensional magnetic iron
  phosphorus trisulfide (FePS3) crystals}},\ }\href
  {https://doi.org/10.1088/2053-1583/3/3/031009} {\bibfield  {journal}
  {\bibinfo  {journal} {2d Mater.}\ }\textbf {\bibinfo {volume} {3}},\ \bibinfo
  {pages} {031009} (\bibinfo {year} {2016})}\BibitemShut {NoStop}%
\bibitem [{\citenamefont {Kim}\ \emph {et~al.}(2019)\citenamefont {Kim},
  \citenamefont {Lim}, \citenamefont {Kim}, \citenamefont {Lee}, \citenamefont
  {Lee}, \citenamefont {Kim}, \citenamefont {Park}, \citenamefont {Son},
  \citenamefont {Park}, \citenamefont {Park},\ and\ \citenamefont
  {Cheong}}]{Kim2019}%
  \BibitemOpen
  \bibfield  {author} {\bibinfo {author} {\bibfnamefont {K.}~\bibnamefont
  {Kim}}, \bibinfo {author} {\bibfnamefont {S.~Y.}\ \bibnamefont {Lim}},
  \bibinfo {author} {\bibfnamefont {J.}~\bibnamefont {Kim}}, \bibinfo {author}
  {\bibfnamefont {J.-U.}\ \bibnamefont {Lee}}, \bibinfo {author} {\bibfnamefont
  {S.}~\bibnamefont {Lee}}, \bibinfo {author} {\bibfnamefont {P.}~\bibnamefont
  {Kim}}, \bibinfo {author} {\bibfnamefont {K.}~\bibnamefont {Park}}, \bibinfo
  {author} {\bibfnamefont {S.}~\bibnamefont {Son}}, \bibinfo {author}
  {\bibfnamefont {C.-H.}\ \bibnamefont {Park}}, \bibinfo {author}
  {\bibfnamefont {J.-G.}\ \bibnamefont {Park}},\ and\ \bibinfo {author}
  {\bibfnamefont {H.}~\bibnamefont {Cheong}},\ }\bibfield  {title} {\bibinfo
  {title} {{Antiferromagnetic ordering in van der Waals 2D magnetic material
  MnPS3 probed by Raman spectroscopy}},\ }\href
  {https://doi.org/10.1088/2053-1583/ab27d5} {\bibfield  {journal} {\bibinfo
  {journal} {2d Mater.}\ }\textbf {\bibinfo {volume} {6}},\ \bibinfo {pages}
  {041001} (\bibinfo {year} {2019})}\BibitemShut {NoStop}%
\bibitem [{\citenamefont {{Le Flem}}\ \emph {et~al.}(1982)\citenamefont {{Le
  Flem}}, \citenamefont {Brec}, \citenamefont {Ouvard}, \citenamefont
  {Louisy},\ and\ \citenamefont {Segransan}}]{LEFLEM1982}%
  \BibitemOpen
  \bibfield  {author} {\bibinfo {author} {\bibfnamefont {G.}~\bibnamefont {{Le
  Flem}}}, \bibinfo {author} {\bibfnamefont {R.}~\bibnamefont {Brec}}, \bibinfo
  {author} {\bibfnamefont {G.}~\bibnamefont {Ouvard}}, \bibinfo {author}
  {\bibfnamefont {A.}~\bibnamefont {Louisy}},\ and\ \bibinfo {author}
  {\bibfnamefont {P.}~\bibnamefont {Segransan}},\ }\bibfield  {title} {\bibinfo
  {title} {{Magnetic interactions in the layer compounds MPX3 (M = Mn, Fe, Ni;
  X = S, Se)}},\ }\href
  {https://doi.org/https://doi.org/10.1016/0022-3697(82)90156-1} {\bibfield
  {journal} {\bibinfo  {journal} {J. Phys. Chem. Solids}\ }\textbf {\bibinfo
  {volume} {43}},\ \bibinfo {pages} {455} (\bibinfo {year} {1982})}\BibitemShut
  {NoStop}%
\bibitem [{\citenamefont {Jiang}\ \emph {et~al.}(2021)\citenamefont {Jiang},
  \citenamefont {Liu}, \citenamefont {Xing}, \citenamefont {Liu}, \citenamefont
  {Guo}, \citenamefont {Liu},\ and\ \citenamefont {Zhao}}]{Jiang2021}%
  \BibitemOpen
  \bibfield  {author} {\bibinfo {author} {\bibfnamefont {X.}~\bibnamefont
  {Jiang}}, \bibinfo {author} {\bibfnamefont {Q.}~\bibnamefont {Liu}}, \bibinfo
  {author} {\bibfnamefont {J.}~\bibnamefont {Xing}}, \bibinfo {author}
  {\bibfnamefont {N.}~\bibnamefont {Liu}}, \bibinfo {author} {\bibfnamefont
  {Y.}~\bibnamefont {Guo}}, \bibinfo {author} {\bibfnamefont {Z.}~\bibnamefont
  {Liu}},\ and\ \bibinfo {author} {\bibfnamefont {J.}~\bibnamefont {Zhao}},\
  }\bibfield  {title} {\bibinfo {title} {{Recent progress on 2D magnets:
  Fundamental mechanism, structural design and modification}},\ }\href
  {https://doi.org/10.1063/5.0039979} {\bibfield  {journal} {\bibinfo
  {journal} {Appl. Phys. Rev.}\ }\textbf {\bibinfo {volume} {8}},\ \bibinfo
  {pages} {031305} (\bibinfo {year} {2021})}\BibitemShut {NoStop}%
\bibitem [{\citenamefont {Li}\ \emph {et~al.}(2020)\citenamefont {Li},
  \citenamefont {Sandhoefner},\ and\ \citenamefont {Kovalev}}]{Li2020}%
  \BibitemOpen
  \bibfield  {author} {\bibinfo {author} {\bibfnamefont {B.}~\bibnamefont
  {Li}}, \bibinfo {author} {\bibfnamefont {S.}~\bibnamefont {Sandhoefner}},\
  and\ \bibinfo {author} {\bibfnamefont {A.~A.}\ \bibnamefont {Kovalev}},\
  }\bibfield  {title} {\bibinfo {title} {{Intrinsic spin Nernst effect of
  magnons in a noncollinear antiferromagnet}},\ }\href
  {https://doi.org/10.1103/PhysRevResearch.2.013079} {\bibfield  {journal}
  {\bibinfo  {journal} {Phys. Rev. Res.}\ }\textbf {\bibinfo {volume} {2}},\
  \bibinfo {pages} {013079} (\bibinfo {year} {2020})}\BibitemShut {NoStop}%
\bibitem [{\citenamefont {Xiao}\ \emph {et~al.}(2005)\citenamefont {Xiao},
  \citenamefont {Shi},\ and\ \citenamefont {Niu}}]{Xiao2005}%
  \BibitemOpen
  \bibfield  {author} {\bibinfo {author} {\bibfnamefont {D.}~\bibnamefont
  {Xiao}}, \bibinfo {author} {\bibfnamefont {J.}~\bibnamefont {Shi}},\ and\
  \bibinfo {author} {\bibfnamefont {Q.}~\bibnamefont {Niu}},\ }\bibfield
  {title} {\bibinfo {title} {Berry phase correction to electron density of
  states in solids},\ }\href {https://doi.org/10.1103/PhysRevLett.95.137204}
  {\bibfield  {journal} {\bibinfo  {journal} {Phys. Rev. Lett.}\ }\textbf
  {\bibinfo {volume} {95}},\ \bibinfo {pages} {137204} (\bibinfo {year}
  {2005})}\BibitemShut {NoStop}%
\bibitem [{\citenamefont {Thonhauser}\ \emph {et~al.}(2005)\citenamefont
  {Thonhauser}, \citenamefont {Ceresoli}, \citenamefont {Vanderbilt},\ and\
  \citenamefont {Resta}}]{Thonhauser2005}%
  \BibitemOpen
  \bibfield  {author} {\bibinfo {author} {\bibfnamefont {T.}~\bibnamefont
  {Thonhauser}}, \bibinfo {author} {\bibfnamefont {D.}~\bibnamefont
  {Ceresoli}}, \bibinfo {author} {\bibfnamefont {D.}~\bibnamefont
  {Vanderbilt}},\ and\ \bibinfo {author} {\bibfnamefont {R.}~\bibnamefont
  {Resta}},\ }\bibfield  {title} {\bibinfo {title} {Orbital magnetization in
  periodic insulators},\ }\href {https://doi.org/10.1103/PhysRevLett.95.137205}
  {\bibfield  {journal} {\bibinfo  {journal} {Phys. Rev. Lett.}\ }\textbf
  {\bibinfo {volume} {95}},\ \bibinfo {pages} {137205} (\bibinfo {year}
  {2005})}\BibitemShut {NoStop}%
\bibitem [{\citenamefont {Shi}\ \emph {et~al.}(2007)\citenamefont {Shi},
  \citenamefont {Vignale}, \citenamefont {Xiao},\ and\ \citenamefont
  {Niu}}]{Shi2007}%
  \BibitemOpen
  \bibfield  {author} {\bibinfo {author} {\bibfnamefont {J.}~\bibnamefont
  {Shi}}, \bibinfo {author} {\bibfnamefont {G.}~\bibnamefont {Vignale}},
  \bibinfo {author} {\bibfnamefont {D.}~\bibnamefont {Xiao}},\ and\ \bibinfo
  {author} {\bibfnamefont {Q.}~\bibnamefont {Niu}},\ }\bibfield  {title}
  {\bibinfo {title} {Quantum theory of orbital magnetization and its
  generalization to interacting systems},\ }\href
  {https://doi.org/10.1103/PhysRevLett.99.197202} {\bibfield  {journal}
  {\bibinfo  {journal} {Phys. Rev. Lett.}\ }\textbf {\bibinfo {volume} {99}},\
  \bibinfo {pages} {197202} (\bibinfo {year} {2007})}\BibitemShut {NoStop}%
\bibitem [{Note1()}]{Note1}%
  \BibitemOpen
  \bibinfo {note} {Note that the orbital moment $\sim \protect \mathbf {r}
  \times \protect \mathbf {v}$ differs from the orbital angular momentum $\sim
  \protect \mathbf {r} \times \protect \mathbf {p}$ by dimension of mass in the
  electronic case~\cite {Bhowal2021}.}\BibitemShut {Stop}%
\bibitem [{\citenamefont {Park}\ and\ \citenamefont {Yang}(2020)}]{Park2020b}%
  \BibitemOpen
  \bibfield  {author} {\bibinfo {author} {\bibfnamefont {S.}~\bibnamefont
  {Park}}\ and\ \bibinfo {author} {\bibfnamefont {B.-J.}\ \bibnamefont
  {Yang}},\ }\bibfield  {title} {\bibinfo {title} {{Phonon Angular Momentum
  Hall Effect}},\ }\href {https://doi.org/10.1021/acs.nanolett.0c03220}
  {\bibfield  {journal} {\bibinfo  {journal} {Nano Lett.}\ }\textbf {\bibinfo
  {volume} {20}},\ \bibinfo {pages} {7694} (\bibinfo {year}
  {2020})}\BibitemShut {NoStop}%
\bibitem [{\citenamefont {Wildes}\ \emph {et~al.}(1998)\citenamefont {Wildes},
  \citenamefont {Roessli}, \citenamefont {Lebech},\ and\ \citenamefont
  {Godfrey}}]{Wildes1998}%
  \BibitemOpen
  \bibfield  {author} {\bibinfo {author} {\bibfnamefont {A.~R.}\ \bibnamefont
  {Wildes}}, \bibinfo {author} {\bibfnamefont {B.}~\bibnamefont {Roessli}},
  \bibinfo {author} {\bibfnamefont {B.}~\bibnamefont {Lebech}},\ and\ \bibinfo
  {author} {\bibfnamefont {K.~W.}\ \bibnamefont {Godfrey}},\ }\bibfield
  {title} {\bibinfo {title} {{Spin waves and the critical behaviour of the
  magnetization in}},\ }\href {https://doi.org/10.1088/0953-8984/10/28/020}
  {\bibfield  {journal} {\bibinfo  {journal} {J. Phys. Condens. Matter}\
  }\textbf {\bibinfo {volume} {10}},\ \bibinfo {pages} {6417} (\bibinfo {year}
  {1998})}\BibitemShut {NoStop}%
\bibitem [{\citenamefont {Sch\"utz}\ \emph {et~al.}(2003)\citenamefont
  {Sch\"utz}, \citenamefont {Kollar},\ and\ \citenamefont
  {Kopietz}}]{Schutz2003}%
  \BibitemOpen
  \bibfield  {author} {\bibinfo {author} {\bibfnamefont {F.}~\bibnamefont
  {Sch\"utz}}, \bibinfo {author} {\bibfnamefont {M.}~\bibnamefont {Kollar}},\
  and\ \bibinfo {author} {\bibfnamefont {P.}~\bibnamefont {Kopietz}},\
  }\bibfield  {title} {\bibinfo {title} {{Persistent Spin Currents in
  Mesoscopic Heisenberg Rings}},\ }\href
  {https://doi.org/10.1103/PhysRevLett.91.017205} {\bibfield  {journal}
  {\bibinfo  {journal} {Phys. Rev. Lett.}\ }\textbf {\bibinfo {volume} {91}},\
  \bibinfo {pages} {017205} (\bibinfo {year} {2003})}\BibitemShut {NoStop}%
\bibitem [{\citenamefont {Sch\"utz}\ \emph {et~al.}(2004)\citenamefont
  {Sch\"utz}, \citenamefont {Kollar},\ and\ \citenamefont
  {Kopietz}}]{Schutz2004PRB}%
  \BibitemOpen
  \bibfield  {author} {\bibinfo {author} {\bibfnamefont {F.}~\bibnamefont
  {Sch\"utz}}, \bibinfo {author} {\bibfnamefont {M.}~\bibnamefont {Kollar}},\
  and\ \bibinfo {author} {\bibfnamefont {P.}~\bibnamefont {Kopietz}},\
  }\bibfield  {title} {\bibinfo {title} {Persistent spin currents in mesoscopic
  haldane-gap spin rings},\ }\href {https://doi.org/10.1103/PhysRevB.69.035313}
  {\bibfield  {journal} {\bibinfo  {journal} {Phys. Rev. B}\ }\textbf {\bibinfo
  {volume} {69}},\ \bibinfo {pages} {035313} (\bibinfo {year}
  {2004})}\BibitemShut {NoStop}%
\bibitem [{\citenamefont {Xia}\ \emph {et~al.}(2023)\citenamefont {Xia},
  \citenamefont {Zhang}, \citenamefont {Liu}, \citenamefont {Zhou},\ and\
  \citenamefont {Ezawa}}]{Xia2023}%
  \BibitemOpen
  \bibfield  {author} {\bibinfo {author} {\bibfnamefont {J.}~\bibnamefont
  {Xia}}, \bibinfo {author} {\bibfnamefont {X.}~\bibnamefont {Zhang}}, \bibinfo
  {author} {\bibfnamefont {X.}~\bibnamefont {Liu}}, \bibinfo {author}
  {\bibfnamefont {Y.}~\bibnamefont {Zhou}},\ and\ \bibinfo {author}
  {\bibfnamefont {M.}~\bibnamefont {Ezawa}},\ }\bibfield  {title} {\bibinfo
  {title} {Universal quantum computation based on nanoscale skyrmion helicity
  qubits in frustrated magnets},\ }\href
  {https://doi.org/10.1103/PhysRevLett.130.106701} {\bibfield  {journal}
  {\bibinfo  {journal} {Phys. Rev. Lett.}\ }\textbf {\bibinfo {volume} {130}},\
  \bibinfo {pages} {106701} (\bibinfo {year} {2023})}\BibitemShut {NoStop}%
\bibitem [{Note2()}]{Note2}%
  \BibitemOpen
  \bibinfo {note} {The electric polarization can be also caused by the Lorentz
  transformation of the magnetic dipole from a moving frame to the laboratory
  ${\protect \bf P} = \protect \frac {{\protect \bf v} \times \protect \bm {\mu
  }}{ c^2} = \protect \frac {g}{2}\protect \frac {e \hbar }{m_{e}
  c^2}({\protect \bf v} \times \protect \hat {{\protect \bf z}})$, where
  ${\protect \bf v}$ and $\protect \bm {\mu } (= g \mu _B {\protect \bf z})$
  are the velocity and the magnetic moment of the magnetic dipole,
  respectively, and $m_{e}$ is the electron mass~\cite {Hirsch1999, Schutz2003,
  MeierPRL2003}. However, this effect is extremely small because the
  denominator $m_{e} c^2$ is order of MeV.}\BibitemShut {Stop}%
\bibitem [{\citenamefont {Choi}\ \emph {et~al.}()\citenamefont {Choi},
  \citenamefont {Jo}, \citenamefont {Ko}, \citenamefont {Go}, \citenamefont
  {Kim}, \citenamefont {Park}, \citenamefont {Kim}, \citenamefont {Min},
  \citenamefont {Choi},\ and\ \citenamefont {Lee}}]{Choi2021}%
  \BibitemOpen
  \bibfield  {author} {\bibinfo {author} {\bibfnamefont {Y.-G.}\ \bibnamefont
  {Choi}}, \bibinfo {author} {\bibfnamefont {D.}~\bibnamefont {Jo}}, \bibinfo
  {author} {\bibfnamefont {K.-H.}\ \bibnamefont {Ko}}, \bibinfo {author}
  {\bibfnamefont {D.}~\bibnamefont {Go}}, \bibinfo {author} {\bibfnamefont
  {K.-H.}\ \bibnamefont {Kim}}, \bibinfo {author} {\bibfnamefont {H.~G.}\
  \bibnamefont {Park}}, \bibinfo {author} {\bibfnamefont {C.}~\bibnamefont
  {Kim}}, \bibinfo {author} {\bibfnamefont {B.-C.}\ \bibnamefont {Min}},
  \bibinfo {author} {\bibfnamefont {G.-M.}\ \bibnamefont {Choi}},\ and\
  \bibinfo {author} {\bibfnamefont {H.-W.}\ \bibnamefont {Lee}},\ }\href@noop
  {} {\bibinfo {title} {{Observation of the orbital Hall effect in a light
  metal Ti}}},\ \Eprint {https://arxiv.org/abs/arXiv:2109.14847}
  {arXiv:2109.14847} \BibitemShut {NoStop}%
\bibitem [{\citenamefont {Han}\ \emph {et~al.}(2022)\citenamefont {Han},
  \citenamefont {Lee},\ and\ \citenamefont {Kim}}]{Han2022}%
  \BibitemOpen
  \bibfield  {author} {\bibinfo {author} {\bibfnamefont {S.}~\bibnamefont
  {Han}}, \bibinfo {author} {\bibfnamefont {H.-W.}\ \bibnamefont {Lee}},\ and\
  \bibinfo {author} {\bibfnamefont {K.-W.}\ \bibnamefont {Kim}},\ }\bibfield
  {title} {\bibinfo {title} {{Orbital Dynamics in Centrosymmetric Systems}},\
  }\href {https://doi.org/10.1103/PhysRevLett.128.176601} {\bibfield  {journal}
  {\bibinfo  {journal} {Phys. Rev. Lett.}\ }\textbf {\bibinfo {volume} {128}},\
  \bibinfo {pages} {176601} (\bibinfo {year} {2022})}\BibitemShut {NoStop}%
\bibitem [{\citenamefont {Cornelissen}\ \emph {et~al.}(2015)\citenamefont
  {Cornelissen}, \citenamefont {Liu}, \citenamefont {Duine}, \citenamefont
  {Youssef},\ and\ \citenamefont {van Wees}}]{Cornelissen2015}%
  \BibitemOpen
  \bibfield  {author} {\bibinfo {author} {\bibfnamefont {L.~J.}\ \bibnamefont
  {Cornelissen}}, \bibinfo {author} {\bibfnamefont {J.}~\bibnamefont {Liu}},
  \bibinfo {author} {\bibfnamefont {R.~A.}\ \bibnamefont {Duine}}, \bibinfo
  {author} {\bibfnamefont {J.~B.}\ \bibnamefont {Youssef}},\ and\ \bibinfo
  {author} {\bibfnamefont {B.~J.}\ \bibnamefont {van Wees}},\ }\bibfield
  {title} {\bibinfo {title} {{Long-distance transport of magnon spin
  information in a magnetic insulator at room temperature}},\ }\href
  {https://doi.org/10.1038/nphys3465} {\bibfield  {journal} {\bibinfo
  {journal} {Nat. Phys.}\ }\textbf {\bibinfo {volume} {11}},\ \bibinfo {pages}
  {1022} (\bibinfo {year} {2015})}\BibitemShut {NoStop}%
\bibitem [{\citenamefont {Goennenwein}\ \emph {et~al.}(2015)\citenamefont
  {Goennenwein}, \citenamefont {Schlitz}, \citenamefont {Pernpeintner},
  \citenamefont {Ganzhorn}, \citenamefont {Althammer}, \citenamefont {Gross},\
  and\ \citenamefont {Huebl}}]{Goennenwein2015}%
  \BibitemOpen
  \bibfield  {author} {\bibinfo {author} {\bibfnamefont {S.~T.~B.}\
  \bibnamefont {Goennenwein}}, \bibinfo {author} {\bibfnamefont
  {R.}~\bibnamefont {Schlitz}}, \bibinfo {author} {\bibfnamefont
  {M.}~\bibnamefont {Pernpeintner}}, \bibinfo {author} {\bibfnamefont
  {K.}~\bibnamefont {Ganzhorn}}, \bibinfo {author} {\bibfnamefont
  {M.}~\bibnamefont {Althammer}}, \bibinfo {author} {\bibfnamefont
  {R.}~\bibnamefont {Gross}},\ and\ \bibinfo {author} {\bibfnamefont
  {H.}~\bibnamefont {Huebl}},\ }\bibfield  {title} {\bibinfo {title}
  {{Non-local magnetoresistance in YIG/Pt nanostructures}},\ }\href
  {https://doi.org/10.1063/1.4935074} {\bibfield  {journal} {\bibinfo
  {journal} {Appl. Phys. Lett.}\ }\textbf {\bibinfo {volume} {107}},\ \bibinfo
  {pages} {172405} (\bibinfo {year} {2015})}\BibitemShut {NoStop}%
\bibitem [{\citenamefont {Hirsch}(1999)}]{Hirsch1999}%
  \BibitemOpen
  \bibfield  {author} {\bibinfo {author} {\bibfnamefont {J.~E.}\ \bibnamefont
  {Hirsch}},\ }\bibfield  {title} {\bibinfo {title} {{Overlooked contribution
  to the Hall effect in ferromagnetic metals}},\ }\href
  {https://doi.org/10.1103/PhysRevB.60.14787} {\bibfield  {journal} {\bibinfo
  {journal} {Phys. Rev. B}\ }\textbf {\bibinfo {volume} {60}},\ \bibinfo
  {pages} {14787} (\bibinfo {year} {1999})}\BibitemShut {NoStop}%
\bibitem [{\citenamefont {Meier}\ and\ \citenamefont
  {Loss}(2003)}]{MeierPRL2003}%
  \BibitemOpen
  \bibfield  {author} {\bibinfo {author} {\bibfnamefont {F.}~\bibnamefont
  {Meier}}\ and\ \bibinfo {author} {\bibfnamefont {D.}~\bibnamefont {Loss}},\
  }\bibfield  {title} {\bibinfo {title} {{Magnetization Transport and Quantized
  Spin Conductance}},\ }\href {https://doi.org/10.1103/PhysRevLett.90.167204}
  {\bibfield  {journal} {\bibinfo  {journal} {Phys. Rev. Lett.}\ }\textbf
  {\bibinfo {volume} {90}},\ \bibinfo {pages} {167204} (\bibinfo {year}
  {2003})}\BibitemShut {NoStop}%
\end{thebibliography}%

\end{document}